\documentclass[a4paper,10pt]{article}
	
\title{On Session Typing, Probabilistic Polynomial Time,\\ and 
Cryptographic Experiments (Long Version)}
\author{Ugo Dal Lago \and Giulia Giusti}
\date{}
	
\usepackage{a4wide}
\usepackage{amsthm}
\theoremstyle{plain}
\newtheorem{theorem}{Theorem}[section]

\newtheorem{lemma}[theorem]{Lemma}
\newtheorem{corollary}{Corollary}
\theoremstyle{definition}
\newtheorem{definition}[theorem]{Definition}

\usepackage{amsmath}
\usepackage{amsfonts}
\usepackage{amssymb}
\usepackage{color}
\usepackage{proof}
\usepackage{mathrsfs}  
\usepackage{bm}
\usepackage{stmaryrd}
\usepackage{mathtools}
\usepackage{bbm}
\usepackage{tikz}
\tikzset{arrow/.style={-{Stealth[]}}}
\usetikzlibrary{positioning,arrows.meta}
\usetikzlibrary{shapes.geometric}
\usepackage{forest}
\usepackage{wrapfig}
\usepackage{empheq}
\usepackage{url}
\usepackage[inference]{semantic}\setpremisesend{0pt}\setpremisesspace{10pt}\setnamespace{0pt}
\usepackage{csquotes}
\usepackage[strings]{underscore}
\usepackage{eucal}
\usepackage{subcaption}
\usepackage{url}
\usepackage{hyperref}
\usepackage{bussproofs}
\usepackage{float}
\usepackage{cleveref}
	
\newcommand{\piDIBLL}{\ensuremath{\pi\mathbf{DIBLL}}}

\newcommand\tab[1][1cm]{\hspace*{#1}}
\newcommand\minitab[1][0.5cm]{\hspace*{#1}}

\newcommand{\spvone}{\mathcal{V}}
\def\mulpol#1{\overline{#1}}
\newcommand{\sbsone}{\rho}
\newcommand{\setpolvar}{\mathcal{PV}}

\newcommand{\distrset}[1]{\mathscr{D}(#1)}
\newcommand{\sd}[1]{\{#1\}}
\newcommand{\sde}[2]{#1^{#2}}
\newcommand{\distrone}{\mathscr{D}}
\newcommand{\distrtwo}{\mathscr{E}}

\newcommand{\indexedprobdistr}[3]{\{#1^{#2}\}_{#3}}
\newcommand{\dirac}[1]{\{#1^1\}}
\newcommand{\multDistr}[2]{#1 \cdot #2}

\newcommand{\sem}[1]{[\![#1]\!]}
\newcommand{\semp}[2]{[\![#1]\!]_{#2}}

\newcommand{\venvone}{\Theta}
\newcommand{\tmone}{a}
\newcommand{\tmtwo}{b}
\newcommand{\tmthree}{c}
\newcommand{\valone}{v}
\newcommand{\valtwo}{w}
\newcommand{\ttrue}{\mathsf{true}}
\newcommand{\tfalse}{\mathsf{false}}
\newcommand{\btone}{B}
\newcommand{\bttwo}{C}
\newcommand{\fsone}{f}
\newcommand{\fstwo}{g}
\newcommand{\funsymbset}{\mathcal{F}}
\newcommand{\NN}{\mathbb{N}}
\newcommand{\typeoffun}[1]{\mathit{typeof}(#1)}
\newcommand{\complof}[1]{\mathit{comof}(#1)}
\newcommand{\algo}[1]{\mathit{algof}(#1)}
\newcommand{\valvarset}{\mathcal{TV}}
\newcommand{\chvarset}{\mathcal{CV}}
\newcommand{\vars}[1]{\mathit{vars}(#1)}
\newcommand{\valred}{\hookrightarrow}
\newcommand{\subst}[2]{\{n\leftarrow p\}}

\newcommand{\procone}{P}
\newcommand{\proctwo}{Q}
\def\parallel#1#2{ #1\mbox{ }|\mbox{ }#2}
\def\res#1#2{ (\nu \mbox{ }#1)\mbox{ }#2}
\def\outchannel#1#2{ #1\langle#2\rangle}
\def\inchannel#1#2{ #1(#2)}
\def\outterm#1#2{ [#1\leftarrow#2]}
\def\letprocess#1#2{\mathtt{let}\mbox{ }#1=#2\mbox{ }\mathtt{in}}
\def\interm#1{ #1}
\def\inrepl#1#2{!#1(#2)}
\def\ifprocess#1#2#3{\mathtt{if} \mbox{ } #1 \mbox{ } \mathtt{then} \mbox{ } #2 \mbox{ } \mathtt{else} \mbox{ } #3}
\def\caseprocess#1#2#3{#1.\mathtt{case}(#2,#3)}
\def\selectfirstprocess#1#2{#1.\mathtt{inl};#2}
\def\selectsecondprocess#1#2{#1.\mathtt{inr};#2}
\newcommand\orsintax{\mbox{ }\bigm|\mbox{ }}
\newcommand{\fn}[1]{\mathit{fn}(#1)}

\newcommand{\subject}[1]{\mathit{s}(#1)}

\def\exptyp#1#2{ !_{#1}#2}
\newcommand\booltyp{\mathbb{B}}
\def\strtyp#1{\mathbb{S}[#1]}

\newcommand{\judgmentobs}[3]{\vdash^{#1}\mbox{ 
		}#2\mbox{ }::\mbox{ }#3}
\newcommand{\judgmentpoly}[6]{#2;\mbox{ }#3;\mbox{ }#4 \mbox{ 
}\vdash^{#1}\mbox{ }#5\mbox{ }::\mbox{ }#6}
\newcommand{\judgmentclosed}[4]{\mathit{#1;\mbox{ }#2\mbox{ }\vdash\mbox{ 
}#3\mbox{ }::\mbox{ }#4}}
\newcommand{\judgmentclosedcorrispondence}[5]{\mathit{#1;\mbox{ }#2\mbox{ }\vdash\mbox{ 
}#3 \rightsquigarrow #4\mbox{ }::\mbox{ }#5}}

\newcommand{\judgmentclosedterm}[5]{\mathit{#2;\mbox{ }#3\mbox{ 
}\vdash^{#1}\mbox{ }#4\mbox{ }::\mbox{ }#5}}
\newcommand{\valtd}[4]{#1\vdash^{#2}#3:#4}
\newcommand{\emptyenv}{\cdot}

\newcommand{\gammapremise}{\Gamma_1 \boxplus \Gamma_2 \sqsubseteq\Gamma}

\def\prooftermsmanipulation#1#2{#1 \equiv\Rightarrow\equiv #2}
\def\expcut#1#2#3#4{cut^{!_{#1}} \mbox{ } (#2)\mbox{ }(#3.#4)}
\def\cut#1#2#3{cut\mbox{ }(#1)\mbox{ }(#2.#3)}

\def\obspredicate#1#2#3#4{ #1\;\downarrow_#3^{#2}\;#4}
\newcommand{\context}{\mathscr{C}[\cdot]}
\newcommand{\contextp}[1]{\mathscr{C}[#1]}
\newcommand{\emptycontext}{[\cdot]}
\def\observeq#1#2{ #1 \cong #2}
\def\contextof#1{ \mathscr{C}[#1]}
\newcommand{\approxdistr}[4]{\ensuremath{#3 \stackrel{\text{#1},\text{#2} }{\approx} #4}}
\newcommand{\setofcontexts}{\mathscr{C}_1[\cdot], \ldots, \mathscr{C}_n[\cdot]}
\def\indexedcontext#1#2{\mathscr{C}_#1[#2]}
\newcommand{\Dlet}{\mathscr{D}_{let}}
\newcommand{\Pletin}{P^{let}_{in}}
\newcommand{\Pinlet}{P^{in}_{let}}
\newcommand{\setofcontextstwo}{\mathscr{B}_1[\cdot], \ldots, \mathscr{B}_m[\cdot]}
\newcommand{\approxdistrlet}[2]{\ensuremath{#1 \stackrel{\text{P}^{let}_{in},\text{P}^{in}_{let} }{\approx} #2}}

\def\eventprob#1{\Pr[#1]}
\def\abs#1{\bigm|#1\bigm|}

\newcommand{\PrivKeavp}[2]{\mathsf{PrivK}^{\mathit{eav}}_{#1,#2}}
\newcommand{\return}[1]{\mathbf{return}\mbox{ }#1}
\newcommand{\inputfrom}[2]{\mathsf{input}\mbox{ } #1 \mbox{ } \mathsf{from} \mbox{ } #2;}
\newcommand{\outputto}[2]{\mathsf{output}\mbox{ } #1 \mbox{ } \mathsf{to} \mbox{ } #2;}

\begin{document}

\maketitle

\begin{abstract}
A system of session types is introduced as induced by a Curry Howard correspondence applied to Bounded Linear 
Logic, then extending the obtained type system with probabilistic 
choice operators and ground types. The resulting system satisfies 
the expected properties, like subject reduction and progress, but also unexpected 
ones, like a polynomial bound on the time needed to reduce processes. 
This  makes the system suitable for modelling experiments and 
proofs from the so-called computational model of cryptography.
\end{abstract}

\section{Introduction}
Session types~\cite{honda1993types, dardha2017session, huttel2016foundations} 
are a typing discipline capable of regulating the interaction between the 
parallel components in a concurrent system in such a way as to \emph{prevent} 
phenomena such as deadlock or livelock, at the same time \emph{enabling} the 
parties to interact following the rules of common communication protocols. In the 
twenty-five years since their introduction, session 
types have been shown to be a flexible tool, being adaptable to 
heterogeneous linguistic and application scenarios (see, 
e.g.,~\cite{SPTD2016,BCDHY2017,ITTVW2019,CHJNY2019}). 
A particularly fruitful line of investigation concerns the links between 
session-type disciplines and Girard's linear logic~\cite{Girard1987}. This 
intimate relationship, known since the introduction of session types, found a 
precise formulation in the work of Caires and Pfenning on a 
Curry-Howard correspondence between session types and intuitionistic linear 
logic~\cite{caires2010session}, which has been developed in multiple 
directions~\cite{toninho2011dependent,wadler2012propositions,perez2014linear,dal2016session}.
 In Caires and Pfenning's type system, proofs of intuitionistic linear logic 
become type derivations for terms of Milner's $\pi$-calculus. Noticeably, 
typable processes satisfy properties (e.g. progress and deadlock freedom) which 
do not hold for untyped processes.

Process algebras, and in particular algebras in the style of the 
$\pi$-calculus, 
have been used, among other things, as specification formalisms for 
cryptographic protocols in the so-called \emph{symbolic} (also known as 
\emph{formal}) model of cryptography, i.e. in the model, due to Dolev and 
Yao~\cite{dolev1983security}, in which aspects related to computational 
complexity and probability theory, themselves central to the 
\emph{computational} model, are abstracted away: strings become symbolic 
expressions, 
adversaries are taken as having arbitrary computing power, and nondeterminism 
replaces probabilism in regulating the interaction between the 
involved parties. This includes $\pi$-calculus dialects akin to the applied 
$\pi$-calculus~\cite{AF2001}, or the spi-calculus~\cite{AG1999}.

Is it possible to model cryptographic protocols by way of process algebras in 
the so-called computational model \emph{itself}? A widely explored path in  
this direction consists in the so-called \emph{computational soundness} results 
for symbolic models, which have been successfully proved in the realm of 
process algebras~\cite{ACF2006,CLHKS2012}. In computationally sound symbolic 
models, any computational attack can be simulated by a symbolic attack, this 
way proving that whenever a protocol is secure in the latter, it must be 
secure in the former, too. If one is interested in calculi precisely and fully 
capturing the computational model, computational soundness is not enough, i.e., 
one wants a model capturing all \emph{and only} the computational adversaries. 
And indeed, there have been some attempts to define process algebras able to 
faithfully capture the computational model by way of operators for 
probabilistic choice and constraints on computational 
complexity~\cite{MRST2006}. The literature, however, is much sparser than for 
process algebras in symbolic style. We believe that this is above all due to 
the fact 
that the contemporary presence of probabilistic evolution and the intrinsic 
nondeterminism of process algebras leads to complex formal systems which are 
hard to reason about.

This paper shows that session typing can be exploited for the sake of designing 
a simple formal system in which, indeed, complexity constraints and 
probabilistic choices can be both taken into account, this way allowing for 
the modelling of cryptographic experiments. At the level of types, we build on 
the approach by Caires and Pfenning, refining it through the lenses of Bounded 
Linear Logic, a logical system which captures polynomial time complexity in the 
sequential setting~\cite{girard1992bounded,HS2004} at the same time allowing 
for a high degree of 
intensional expressivity~\cite{DLH2010}. At the level of processes, we enrich 
proof 
terms with first-order function symbols computing probabilistic polytime 
functions, namely the basic building blocks of any cryptographic protocol. This 
has two consequences: process evolution becomes genuinely probabilistic, while 
process terms and types are enriched so as to allow for the exchange of 
strings, this way turning the calculus to an applied one. From a purely 
definitional perspective, then, the introduced calculus, called $\piDIBLL$, is 
relatively simple, and does not significantly deviate from the literature, 
being obtained by mixing well-known ingredients in a novel way. 
The calculus $\piDIBLL$ is introduced in 
Section~\ref{sect:processesandsessions} below.

Despite its simplicity $\piDIBLL$ is on the one hand capable of 
expressing some simple cryptographic experiments, and on the other hand 
satisfies 
some strong meta-theoretical properties. This includes type soundness, which is 
expected, and can be spelled out as \emph{subject reduction} and \emph{progress}, but also a 
polynomial bound on the length of reduction sequences, a form of 
\emph{reachability} property which is essential for our calculus to be 
considered a model of cryptographic adversaries. All this is described in 
Section~\ref{sect:soundessandpoly}. 

As interesting as they are, these properties are not by themselves sufficient 
for considering $\piDIBLL$ a proper calculus for computational cryptography. 
What is missing, in fact, is a way to capture computational 
indistinguishability, in the sense of the computational 
model~\cite{katz2020introduction,Goldreich2006}. Actually, this is where the 
introduced calculus shows its peculiarities with respect to similar calculi 
from the literature, and in particular with respect to the CCS-style calculus 
by Mitchell and Scedrov~\cite{MRST2006}. Indeed, $\piDIBLL$ 
typable processes enjoy a confluence property which cannot hold for untyped 
processes. The latter, in turn, implies that firing internal actions on any 
typable process results in a \emph{unique} distribution of processes, all of 
them ready to  produce an observable action. This makes relational reasoning 
handier. We in 
particular explore observational equivalence in Section~\ref{sect:confluence}, 
then showing how this  can be of help in a simple experiment-based security 
proof in Section~\ref{sect:cryptoproof}.

\section{A Bird's Eye View on Cryptographic Experiments and 
Sessions}\label{sect:birdeyeview}

\newcommand{\Gen}{\mathit{Gen}}
\newcommand{\Enc}{\mathit{Enc}}
\newcommand{\Dec}{\mathit{Dec}}
\newcommand{\esone}{\Pi}
\newcommand{\PrivK}{\mathsf{PrivK}^{\mathit{eav}}}
\newcommand{\ProcPrivK}{\mathit{PRIVK}}
\newcommand{\advc}{\mathit{adv}}
\newcommand{\expc}{\mathit{exp}}
\newcommand{\schc}{\mathit{sch}}
\newcommand{\ProcAdv}{\mathit{ADV}}
In this section, we introduce the reader to cryptographic experiments, and we 
show how they and the parties involved can be conveniently modelled as 
session-typed processes. We will also hint at how relational reasoning could be 
useful in supporting proofs of security. We will do all this by way of an 
example, namely the one of private key encryption schemes and security against 
passive adversaries. We will try to stay self-contained, and the interested 
reader can check textbooks~\cite{katz2020introduction} for more details or for the necessary 
cryptographic preliminaries. As examples we recall the following notions
\begin{definition}[Negligible Function]
A function $f$ from the natural numbers to the non-negative real numbers is negligible iif for every positive polynomial $p$ there is an $N\in \mathbb{N}$ such that for all natural numbers $n>N$ it holds that $f(n)<1/p(n)$.
\end{definition}

\begin{definition}[Probabilistic Polynomial Time (PPT) Algorithm]
A probabilistic algorithm $A$ is called PPT iff there exists a polynomial $p$ which is an upper limit to the computational complexity of $A$ regardless of the probabilistic choices made by the latter.
\end{definition}
Since the running time of any cryptographic algorithm has to be polynomially bounded w.r.t. the \emph{value} of the security parameter $n$, the latter is passed in unary (i.e. as $1^n$) to the algorithm, so that $n$ is also a lower bound to the \emph{length} of the input.

\begin{figure}
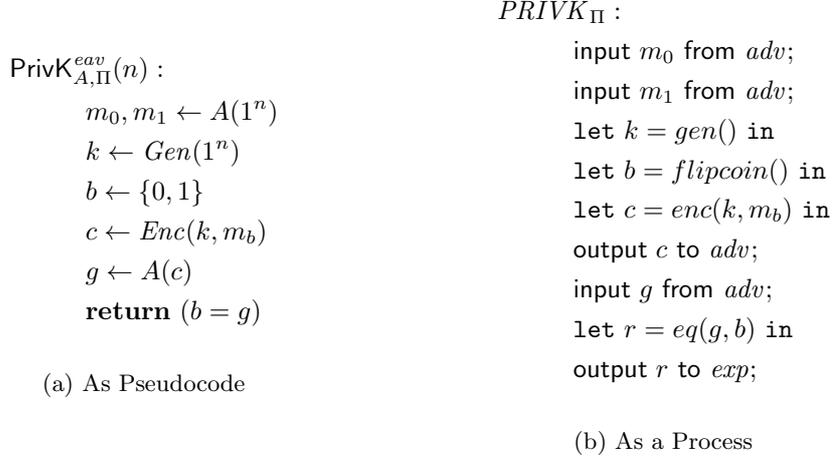

	\centering
	\begin{subfigure}[c]{.45\linewidth}
		\centering
		\begin{align*}
		    &\PrivKeavp{A}{\Pi}(n):\\
		    &\tab m_0, m_1 \leftarrow A(1^n)\\
		    &\tab k \leftarrow \Gen(1^n)\\
		    &\tab b \leftarrow \{0,1\}\\
		    &\tab c \leftarrow \Enc(k,m_b)\\
		    &\tab g \leftarrow A(c)\\
		    &\tab \return{(b=g)}
		\end{align*}
		\caption{As Pseudocode}\label{fig:indistinguishpseudo}
	\end{subfigure}
	\begin{subfigure}[c]{.45\linewidth}
		\centering
		\begin{align*}
		    &\ProcPrivK_\esone:\\
		    &\tab \inputfrom{m_0}{\advc}\\
		    &\tab \inputfrom{m_1}{\advc}\\
		    &\tab \letprocess{k}{gen()}\\
		    &\tab \letprocess{b}{flipcoin()}\\
		    &\tab \letprocess{c}{enc(k,m_b)}\\
		    &\tab \outputto{c}{\advc}\\
		    &\tab \inputfrom{g}{\advc}\\
		    &\tab \letprocess{r}{eq(g,b)}\\
		    &\tab \outputto{r}{\expc}
		\end{align*}
		\caption{As a Process}\label{fig:indistinguishprocess}
	\end{subfigure}
	\caption{The Indistinguishability Experiment}\label{fig:indistinguish}
\end{figure}

A private-key encryption scheme is a triple of algorithms 
$\esone=(\Gen,\Enc,\Dec)$, the first one responsible for key generation, the 
latter two being the encryption and decryption algorithms, respectively. When 
could we say that such a scheme $\esone$ is \emph{secure}? Among the many 
equivalent definitions in the literature, one of the handiest is the one based 
on indistinguishability, which is based on the experiment $\PrivK$ reported in 
Figure~\ref{fig:indistinguishpseudo} exactly in the form it has
in~\cite{katz2020introduction}.
As the reader may easily notice, the experiment is nothing more than a 
randomized algorithm interacting with both the adversary $A$ and the scheme 
$\Pi$. The interaction between $\PrivK$ and the adversary $A$ can be put in 
evidence by switching to a language for processes, see 
Figure~\ref{fig:indistinguishprocess}. The process $\ProcPrivK_\esone$ 
communicates with the adversary through the channel $\advc$ and outputs the 
result of its execution to the channel $\expc$. Apart from the fact that the 
adversary has been factored out, the process is syntactically very similar to 
the experiment $\PrivK$. Actually, we could have made the 
interaction between $\ProcPrivK$ and $\esone$ explicit by turning the latter 
into a process interacting with the former through a dedicated channel.

The interaction between an adversary $\ProcAdv$ and $\ProcPrivK_\Pi$ can be 
modelled through the parallel composition operator, i.e., by studying the 
behaviour of $\parallel{\ProcAdv}{\ProcPrivK_\Pi}$. As we will soon see, we 
would like the aforementioned parallel composition to output $\ttrue$ on the 
channel $\expc$ with probability very close to $\frac{1}{2}$, and this is 
indeed what cryptography actually prescribes~\cite{katz2020introduction}. We 
should 
not, however, be too quick to proclaim the problem solved. What, for example, 
if $\ProcAdv$ communicates with $\ProcPrivK_\Pi$ in a way different from the 
one prescribed by the experiment, e.g. by \emph{not} passing two strings to it, 
thus blocking the interaction? Even worse, what if $\ProcPrivK_\esone$ becomes 
the parallel composition $\parallel{\ProcPrivK}{\Pi}$ and $\ProcAdv$ cheats on 
the communication by intercepting the messages exchanged between $\Pi$ and 
$\ProcPrivK$? These scenarios are of course very interesting from a security 
viewpoint, but we are not interested at those here: the only thing $\ProcAdv$ 
is allowed to do is to send the two messages and to use its internal 
computational capabilities to guess the value $b$ the experiment produces.

How to enforce all this at the level of processes? Actually, this is what 
session types are good for! It would be nice, for example, to be able to type 
the two processes above as follows:
\begin{align*}
\advc:\strtyp{p} \otimes \strtyp{p} \otimes (\strtyp{p} \multimap 
\booltyp)\vdash&\ProcPrivK_\Pi::\expc:\booltyp\\
\vdash&\ProcAdv::\advc:\strtyp{p} \otimes \strtyp{p} \otimes (\strtyp{p} 
\multimap \booltyp)
\end{align*}
where $\booltyp$ is the type of booleans and $\strtyp{p}$ is the type of 
strings of length $p$.
Moreover, we would like to somehow force a restriction $\nu\advc$ to be placed 
next to the parallel composition $\parallel{\ProcAdv}{\ProcPrivK_\Pi}$, so as 
to prescribe that $\ProcAdv$ can only communicate with the experiment, and not 
with the outside world. Finally, we would like $\ProcAdv$ to range over 
processes working in polynomial time. All 
this is indeed taken care of by our session type discipline as introduced in 
Section~\ref{sect:processesandsessions}.

But now, would it be possible to not only \emph{express} simple cryptographic 
situations, but also to \emph{prove} some security properties about them from 
within the realm of processes? As already mentioned, this amounts to requiring 
that for every efficient adversary (i.e. for every PPT algorithm) $A$ it holds 
that $\Pr[\PrivKeavp{A}{\Pi}(n)]\leq\frac{1}{2}+\varepsilon(n)$, where 
$\varepsilon$ is negligible. In the realm of processes, this becomes the 
following equation
\begin{equation}\label{eq:security}
	\nu\advc.(\parallel{\ProcPrivK_\Pi}{\ProcAdv})\sim\mathit{FAIRFLIP}_{\expc}
\end{equation}
where $\mathit{FAIRFLIP}_{\expc}$ behaves like a fair coin outputting its value 
on the channel $\expc$, and $\sim$ expresses approximate equivalence as induced 
by negligible functions. Making all this formal is nontrivial for at least 
three reasons:
\begin{itemize}
\item
  First of all, the statement only holds for \emph{efficient} adversaries. The 
  relation $\sim$, however specified, must 
  then take this constraint into account.
\item
  Secondly, the relation $\sim$ only holds in an approximate sense, and the 
  acceptable degree of approximation crucially depends on $n$, the so-called 
  security parameter. This is due to negligibly, without which cryptography 
  would be essentially vacuous.
\item
  Finally, the computational security of $\Pi$ can at the time of writing be 
  proved only based on \emph{assumptions}, e.g. that one-way functions or 
  pseudorandom 
  generators exist. In other words, Equation~(\ref{eq:security}) only holds in 
  a conditional sense, and cryptographic proofs have to be structured 
  accordingly.
\end{itemize}
The calculus $\piDIBLL$ successfully addresses all these challenges, as we are 
going to show in the rest of this paper.

\section{Processes and Session Typing}\label{sect:processesandsessions}
\newcommand{\piDILL}{\pi\mathbf{DILL}}
This section is devoted to introducing $\piDIBLL$, a variation on $\piDILL$~\cite{caires2010session} in which 
a polynomial constraint on the replicated processes is enforced following the 
principles of Bounded Linear Logic~\cite{girard1992bounded}. For the sake of 
properly representing cryptographic protocols in the computational model, 
$\piDIBLL$ is also equipped with indexed ground types and a notion of 
probabilistic choice.

\subsection{Preliminaries}
Preliminary to the definition of the $\piDIBLL$ session type system are three 
concepts, namely polynomials, probability distributions and indexed ground 
types. Let us start this section introducing polynomials.

\begin{definition}[Polynomials]
\emph{Polynomial variables} are indicated with metavariables like $n$ and $m$, 
and form a set $\setpolvar$. \emph{Polynomials expressions} are built from 
natural number constants, polynomial variables, addition and multiplication. A 
polynomial $p$ depending on the polynomial variables $\mulpol{n}=n_1,\dots,n_k$ 
is sometime indicated as $p(n_1,\ldots,n_k)$ and abbreviated as 
$p(\mulpol{n})$. Such a polynomial is said to be a \emph{$\spvone$-polynomial} 
whenever all variables in the sequence $\mulpol{n}$ are in 
$\spvone\subseteq\setpolvar$. If $\spvone\subseteq\setpolvar$, any map 
$\sbsone:\spvone\rightarrow\NN$ is said to be a \emph{$\spvone$-substitution}, 
and the natural number obtained by interpreting any variable $m\in\spvone$ 
occurring in a $\spvone$-polynomial $p$ with $\sbsone(m)$ is indicated just as 
$p(\sbsone)$. If $\spvone$ is a singleton $\{n\}$, the substitution mapping $n$ 
the to the natural number $i$ is indicated as 
$\sbsone_i$, and $\spvone$ is indicated, abusing notation, with $n$. 
The set of polynomial variables occurring in any polynomial $p$ is 
indicated as $\vars{p}$, and this definition will be silently extended to other 
kinds of syntactic objects like terms and types.
\end{definition}
Distributions play a crucial role in probability theory and represent the 
likelihood of observing an element from given set. In this paper, they will be 
the key ingredient in giving semantics to types and processes.
\begin{definition}[Probability Distributions] \label{probDistr}
A \emph{probability distribution} on the finite set $A$ is a function 
$\mathcal{D}: A 
\rightarrow \mathbb{R}_{[0,1]}$ such that: $\sum_{v\in A} \mathcal{D}(v) =1$. A 
probability distribution is often indicated by way of the the notation 
$\sd{\sde{v_1}{r_1},\ldots,\sde{v_m}{r_m}}$ (where $v_1,\ldots,v_m$ are 
distinct elements of $A$), which stands for the 
distribution $\mathcal{D}$ such that $r_i = 
\mathcal{D}(v_i)$ for every $i \in \{1,\ldots,m\}$.
Given a probability distribution $\mathcal{D}$ on $A$, its \emph{support}
$S(\mathcal{D})\subseteq A$ contains precisely those elements of $A$ to which 
$\mathcal{D}$ attributes a strictly positive probability. The set of all 
probability distributions on a set $A$ is indicated as $\distrset{A}$.
\end{definition}
In the computational model, the agents involved exchange binary strings. We 
keep the set of ground types slightly more general, so as to treat booleans as 
a separate type. As a crucial step towards dealing with polytime 
constraints, the type of strings is indexed by a polynomial, which captures the 
length of binary strings inhabiting the type.
\begin{definition}[Ground Types] \label{baseTypes}
\emph{Ground types} are expressions generated by the grammar
$\btone::=\booltyp\orsintax\strtyp{p}$, 
where $p$ is a polynomial expression. A $\spvone$-ground type is a ground type 
$\btone$ such that all polynomial variables occurring in it are taken from 
$\spvone$, and as such can be given a semantics in the context of a 
$\spvone$-substitution $\sbsone$: $\semp{\booltyp}{\sbsone}=\{0,1\}$ , 
$\semp{\strtyp{p}}{\sbsone}=\{0,1\}^{p(\sbsone)}$.
\end{definition}
The concrete nature of any ground type $\strtyp{p(\mulpol{n})}$ is only known 
when the polynomial variables in $\mulpol{n}$, which stand for so-called 
security parameters, are attributed a natural number value. For reasons of 
generality, we actually allow more than one security parameter, even if 
cryptographic constructions almost invariably need only one of them.

\subsection{Terms}
Terms are expressions which are \emph{internally} evaluated by processes, the 
result of this evaluation having a ground type and being exchanged between 
the different (sub)processes. 
\subparagraph*{Function Symbols.}
We work with a set $\funsymbset$ of \emph{function symbols}, ranged over by 
metavariables like $\fsone$ and $\fstwo$. In the context of this paper, it is 
important that function symbols can be evaluated in probabilistic polynomial 
time, and this can be achieved by taking function symbols from a language 
guaranteeing the aforementioned complexity bounds~\cite{MMS1998,DLZG2014}. Each 
function symbol $\fsone\in\funsymbset$ comes equipped with:
\begin{itemize}
    \item
      A type $\typeoffun{\fsone}$ having the form 
      $\btone_1,\ldots,\btone_m\rightarrow\bttwo$ where the $\btone_i$ and 
      $\bttwo$ are $\{n\}$-ground types.
    \item
      A family $\sem{\fsone}=\{\semp{\fsone}{i}\}_{i\in\NN}$ of functions giving semantics to $\fsone$ such that 
      $\semp{\fsone}{i}$ goes from 
      $\semp{\btone_1}{\sbsone_i}\times\ldots\times\semp{\btone_m}{\sbsone_i}$
      to $\distrset{\semp{\bttwo}{\sbsone_i}}$,
      where $\typeoffun{\fsone}$ is $\btone_1,\ldots,\btone_m\rightarrow\bttwo$.
    \item We assume each function symbol $\fsone$ to be associated with an 
    $n$-polynomial $\complof{\fsone}$ bounding the complexity of computing 
    $\fsone$, in the following sense: there must be a PPT algorithm 
    $\algo{\fsone}$ which, on input $1^i$ and a tuple $t$ in 
    $\semp{\btone_1}{\sbsone_i}\times\cdots\times\semp{\btone_m}{\sbsone_i}$
    returns in time at most $\complof{\fsone}(\rho_i)$ each value $x\in\semp{\bttwo}{\sbsone_i}$ with probability $\semp{\fsone}{i}(t)(x)$, where $\typeoffun{\fsone}=\btone_1,\ldots,\btone_m\rightarrow\bttwo$.
\end{itemize}
\subparagraph*{Term Syntax and Semantics.}
Finally, we are able to define terms and values, which are expressions 
derivable in the following grammars:
\begin{align*}
\tmone,\tmtwo,\tmthree&::= \valone \orsintax 
\fsone_p(\valone_1,\ldots,\valone_n) & 
\valone,\valtwo&::= z \orsintax  \ttrue \orsintax  \tfalse \orsintax  s. & 
\end{align*}
Here $\fsone$ is a function symbol, $p$ is a polynomial,
$\ttrue$ and $\tfalse$ are the usual boolean constants, $s$ is any binary 
string, and
$z$ is a \emph{term variable} taken from a set $\valvarset$ disjoint from 
$\setpolvar$. Terms are assumed to be well-typed according to an elementary 
type system which will be defined later.
Reduction rules between terms and distributions of values are given 
only for terms which are closed with respect to both term variables and 
polynomial variables. Here are the rules
\begin{center}
	\AxiomC{}
	\UnaryInfC{$\valone\valred\dirac{\valone}$}
	\DisplayProof
	\quad\quad
	\AxiomC{}
	\UnaryInfC{$\fsone_i(\valone_1,\ldots,\valone_m)\valred\semp{\fsone}{i}(\valone_1,\ldots,\valone_m)$}
    	\DisplayProof
\end{center}

\subsection{Processes}
It is finally time to introduce the process terms of $\piDIBLL$, which as 
already  mentioned are a natural generalization of those of $\piDILL$~\cite{caires2010session}.

\begin{definition}[Process Syntax]
Given an infinite set of names, the set of processes, indicated with 
metavariables like $\procone$ and $\proctwo$ is defined by the following 
grammar:
\begin{align*}
P,Q ::= \mbox{ } & 0 \orsintax \parallel{P}{Q} \orsintax \res{y}{P} \orsintax \outchannel{x}{y}.P \orsintax \inchannel{x}{y}.P  \orsintax \outterm{x}{v} \orsintax \letprocess{x}{a}\mbox{ }P  \orsintax \\ 
&  \interm{x}.P \orsintax \inrepl{x}{y}.P \orsintax \selectfirstprocess{x}{P} \orsintax \selectsecondprocess{x}{P} \orsintax \caseprocess{x}{P}{Q} \orsintax \ifprocess{\valone}{P}{Q}
\end{align*}
where $\mathit{x,y}$ are channel names from a set $\chvarset$ such that 
$\valvarset\subseteq\chvarset$, $\tmone$ is a term and $\valone$ is a 
value. 
\end{definition}
The operators used in the process syntax have the following meaning:
\begin{itemize}
	\item The process $\mathit{0}$ is the \emph{inactive process}, that is a terminated process.
	\item The process $\mathit{\parallel{P}{Q}}$ is the \emph{parallel composition} between two processes, in which $\procone$ and $\proctwo$ interact 
	with 	each other and with the context.
	\item The \emph{name restriction} operator is used to make the name of a process private. In other words, $\mathit{\res{y}{P}} $ can be seen as the 		process that assigns a new name (different from any names possibly used by any other process in the context) and proceeds according to $\procone$.
	\item $\mathit{\outchannel{x}{y}.P}$ is the \emph{output process} that sends the channel $y$ on $x$ and then proceeds according to $\procone$.
	\item $\mathit{\inchannel{x}{y}.P}$ is the \emph{input process} that receives a channel $z$ on  $x$  and then proceeds according to $\procone$ 
	where the 	name $y$ is replaced by $z$.
	\item $\mathit{\outterm{x}{v}}$ is the \emph{output process of a value} responsible for sending to the channel $x$ the value $\valone$.
	\item $\mathit{\interm{x}.P}$ is the \emph{value input process} and it receives a value through the channel $x$, which is then substituted in the 
	process $\procone$.
	\item $\mathit{\letprocess{x}{a}\mbox{ }P}$ is the \emph{term's evaluation process} where channel $x$ in the process $P$ assumes the values 
	obtained from the evaluation of the term $\tmone$.
	\item $\mathit{\inrepl{x}{y}.P}$ is the \emph{replicated process or persistent process}. It performs the input operation an arbitrary number of times.
	\item The \emph{binary choice operator} provides two labels in which the first involves the execution of the process $\procone$, whereas the second 			executes the process $\proctwo$. The $\mathit {\caseprocess{x}{P}{Q}}$ process gives the choice between two processes denoted by $\procone$ and 	$\proctwo$.
	\item The process $\mathit{\selectfirstprocess{x}{P}}$ is the process that selects the first label provided by a binary choice operation and then 
	proceeds according to $\procone$.
	\item The process $\mathit{\selectsecondprocess{x}{P}}$ is the process that selects the second label provided by a binary choice operation and then 			proceeds according to $\procone$.
	\item The process $\mathit{\ifprocess{\valone}{P}{Q}}$ is a \emph{conditional}. Observe that the value $\valone$ can be either $\ttrue$, $\tfalse$, or 
	a variable.
\end{itemize}
For any process $P$, we denote the set of free names of $P$ by $\fn{\procone}$ which is defined by induction (on the process syntax) as follows:
\allowdisplaybreaks
\begin{align*}
	fn(x) &= \{x\}\\
	fn(0) &= \emptyset\\
	fn(\parallel{P}{Q}) &= fn(P) \cup fn(Q)\\
	fn(\res{y}{P}) &=  fn(P) \setminus\{y\}\\
	fn(\outchannel{x}{y}.P) &= fn(x) \cup fn(y) \cup fn(P)\\
	fn(\inchannel{x}{y}.P) &= fn(x) \cup fn(P) \setminus \{y\}\\
	fn(\outterm{x}{v}) &= fn(x)\\
	fn(\letprocess{x}{a} \mbox{ }P) &= fn(x) \cup fn(P)\\
	fn(\interm{x}.P) &= fn(P) \setminus \{x\}\\
	fn(\inrepl{x}{y}.P) &= fn(x) \cup fn(P) \setminus \{y\}\\
	fn(\selectfirstprocess{x}{P}) &= fn(x) \cup fn(P)\\
	fn(\selectsecondprocess{x}{P}) &= fn(x) \cup fn(P)\\
	fn(\caseprocess{x}{P}{Q}) &= fn(x) \cup fn(P) \cup fn(Q)\\
	fn(\ifprocess{\valone}{P}{Q}) &= fn(P) \cup fn(Q)
\end{align*}
The grammar of processes we have just introduced is perfectly adequate to 
represent the process $\ProcPrivK_\Pi$ as from 
Figure~\ref{fig:indistinguishprocess}.

\subsection{Process Reduction}\label{sect:reduction}

Process reduction in $\piDIBLL$ is intrinsically probabilistic, and as such 
deserves to be described with some care. 

\begin{definition}[Structural Congruence]
Structural congruence $\equiv$ is the least congruence on processes satisfying 
the following axioms:

\begin{align*}
	\res{x}{(\parallel{P}{\res{y}{(\parallel{Q}{R})}})} & \equiv \res{y}{(\parallel{\res{x}{(\parallel{P}{Q})}}{R})} \quad 	&&x \notin \fn{R} \land y \notin \fn{P}\\
	\res{x}{(\parallel{P}{\res{y}{(\parallel{Q}{R})}})} & \equiv \res{y}{(\parallel{Q}{\res{x}{(\parallel{P}{R})}})} \quad 	&&x \notin \fn{Q} \land y \notin \fn{P}\\
	P\equiv_\alpha Q \mbox{ } \Rightarrow \mbox{ } P&\equiv Q\\
	P &\equiv \res{x}{(\parallel{P}{0})} \quad &&x \notin \fn{P}
\end{align*}
\end{definition}
The reduction relation between processes is \emph{not} a plain binary relation 
anymore, and instead puts a process $\procone$ in correspondence with a 
\emph{distribution} $\distrone$ of processes, namely an object in the form 
$\sd{\sde{\procone_1}{r_1},\ldots,\sde{\procone_m}{r_m}}$, where the 
$\procone_i$ are processes and the $r_i$ are positive real numbers summing to 
$1$. We write $\procone\rightarrow\distrone$ in this case.

The reduction rules for processes are defined as follows:
\begin{center}
	\AxiomC{}
    	\UnaryInfC{$\parallel{\outchannel{x}{y}.Q}{\inchannel{x}{u}.P}\rightarrow \dirac{\parallel{Q}{P\{y/u\}}}$}
	\DisplayProof
	\quad\quad
	\AxiomC{}
    	\UnaryInfC{$\parallel{\outchannel{x}{y}.Q}{\inrepl{x}{u}.P}\rightarrow \dirac{\parallel{Q}{\parallel{P\{y/u\}}{\inrepl{x}{u}.P}}}$}
    	\DisplayProof
\end{center}

\begin{center}
	\AxiomC{}
    	\UnaryInfC{$\parallel{\selectfirstprocess{x}{P}}{\caseprocess{x}{Q}{R}} \rightarrow \dirac{\parallel{P}{Q}}$}
	\DisplayProof
	\quad\quad
	\AxiomC{}
    	\UnaryInfC{$\parallel{\selectsecondprocess{x}{P}}{\caseprocess{x}{Q}{R}} \rightarrow \dirac{\parallel{P}{R}}$}
    	\DisplayProof
\end{center}

\begin{center}
	\AxiomC{$Q \rightarrow \indexedprobdistr{Q_i}{r_i}{i\in I}$}
	\UnaryInfC{$\parallel{P}{Q}\rightarrow \indexedprobdistr{\parallel{P}{Q_i}}{r_i}{i\in I}$}
	\DisplayProof
	\quad\quad
	\AxiomC{$P\rightarrow \indexedprobdistr{Q_i}{r_i}{i \in I}$}
	\UnaryInfC{$\res{y}{P}\rightarrow \indexedprobdistr{\res{y}{Q_i}}{r_i}{i\in I}$}
    	\DisplayProof
\end{center}

\begin{prooftree}
\AxiomC{$P\equiv P'$}
\AxiomC{$P' \rightarrow \indexedprobdistr{Q_i}{r_i}{i \in I}$}
\AxiomC{$Q_i \equiv Q'_i$ \textit{where} $i\in I$}
\TrinaryInfC{$P \rightarrow  \{Q_i^{\mbox{ } r_i}\}_{i\in I}$}
\end{prooftree}

\begin{center}
	\AxiomC{$\valone\valred \dirac{val}$}
	\UnaryInfC{$\parallel{\outterm{x}{v}}{\interm{x}.Q}\rightarrow \dirac{Q\{val /x \}}$}
	\DisplayProof
	\quad\quad
	\AxiomC{$a \valred  \indexedprobdistr{v_i}{r_i}{i \in I}$}
	\UnaryInfC{$\letprocess{x}{a}\mbox{ }P \rightarrow \indexedprobdistr{P\{v_i/x\}}{r_i}{i \in I}$}
    	\DisplayProof
\end{center}

\begin{center}
	\AxiomC{}
    	\UnaryInfC{$\ifprocess{\ttrue}{P}{Q} \rightarrow \dirac{P}$}
	\DisplayProof
	\quad\quad
	\AxiomC{}
    	\UnaryInfC{$\ifprocess{\tfalse}{P}{Q} \rightarrow \dirac{Q}$}
    	\DisplayProof
\end{center}

\subsection{Type System}

Traditionally, session typing serves the purpose of guaranteeing safety 
properties, like the absence of deadlocks. In this paper, however, they also 
enforce some bounds on the complexity of the reduction process, and 
as such have to be made more restricted.

\subparagraph*{Types.}
First of all, let us introduce the language of types, which is defined as 
follows:
\begin{center}
	$A,B\mbox{ } ::=\mbox{ }1\orsintax  A\multimap B \orsintax \exptyp{p}{A}\orsintax  A \otimes B\orsintax  A \oplus B \orsintax  A \& B \orsintax  \booltyp 			\orsintax \strtyp{p}$
\end{center}
where:
\begin{itemize}
	\item $1$ is the type of an empty or terminated session channel. A process offering to communicate via a session channel typed this way simply 					synchronizes with another process through it without exchanging anything.
	\item $A\multimap B$ is the type of a session channel $x$ through which a message carrying another channel with type $A$ is received. After 			
	performing this action, the underlying process behaves according to $B$ on the same channel $x$.
	\item $!_{p}A$ is the type for a replicated process which can be used as a server to generate a limited number $p$ of new sessions with type $A$. In 
	other words, $!_{p}A$ is the type of a process which offers to open $p$ new sessions of type $A$, where $p$ is a $\spvone$-polynomial used to limit 
	the number of copies of a replicated process.
	\item $A \otimes B$ is the type of a session channel $x$ through which a message carrying another channel with type $A$ is sent. After performing this 		action, the underlying process behaves according to $B$ on the same channel $x$.
	\item $A \oplus B$ is the type of a selection session. More precisely, it is the type of a channel on which a process either sends a special message $inl$ 	
	and performs according to $A$ or sends a special message $inr$ and performs according to $B$. This corresponds to an internal choice.
	\item The type $\mathit{A \& B}$ can be assigned to a channel $x$ on which the underlying process offers the possibility of choosing between 
	proceeding according to $A$ or to $B$, both on $x$. This corresponds to an external choice.
	\item $\booltyp$ is the base type for boolean values denoted by: $true$ and $false$.
	\item $\strtyp{p}$ is the base type for binary strings with polynomial length.
\end{itemize}

\subparagraph*{Type Environments.}
 The \emph{type environment} is divided into the following three parts: 
\begin{itemize}
	\item $\Delta$ is the linear part of the type environment, it contains assignments $x: A$ where $x$ is the name of a channel typed by $A$. These channels can be used only once.
	Formally, $\Delta$ is defined by induction as follows
	$$\Delta := \emptyenv \orsintax \Delta, \mbox{ } x:A $$
	where $x$ is the name of a linear channel, and $A$ is a type.
	
	\item The unrestricted part of the type environment, denoted by $\Gamma$, contains assignments $x_{p}: A$ where $x$ is the name of an unrestricted 	channel indexed by the polynomial $p$ and typed by $A$. Such channels can be used a polynomial number of times, this limitation is denoted by $p$.  			Formally, $\Gamma$ is defined by induction as follows
	    $$\Gamma := \emptyenv \orsintax \Gamma, \mbox{ } x_{p}:A$$
	where $x$ is the name of an unrestricted channel which can be used a maximum of $p$ times, and $A$ is a session type.

	\item The third and last part $\Theta$ contains assignments $x:A$ where $x$ is a variable for terms and $A$ is the base type associated with that 		
	variable. Formally, $\Theta$ is defined by induction as follows
    	$$ \Theta := \emptyenv \orsintax \Theta, \mbox{ } x:A$$
    where $x$ is a variable for terms, and $A$ is a ground type.
\end{itemize}

Unrestricted environment have to be manipulated with great care while typing 
processes, in particular in all binary typing rules. This requires the 
introduction of a (partial) binary operation $\boxplus$ on unrestricted type 
environments and a partial order relation $\sqsubseteq$, which are defined as follows:
\begin{definition}[$\boxplus$ Operation] \label{envSumOperation} The $\boxplus$ operation  takes as input two unrestricted type environments, $\Gamma_1$ and $\Gamma_2$, and outputs an unrestricted type environment $\Gamma$ such that: for all unrestricted channels $x_{s(\mulpol{n})} \in \Gamma$ there exists two polynomials $q(\mulpol{n})$ and $r(\mulpol{n})$ so that $x_{q(\mulpol{n})} \in \Gamma_1$ and $x_{r(\mulpol{n})} \in \Gamma_2$, and these polynomials are such that $s(\mulpol{n}) = q(\mulpol{n})+r(\mulpol{n})$.
\end{definition}

\begin{definition}[$\sqsubseteq$ Relation] \label{envPartialOrdRel} Given two unrestricted type environment \\ $\mathit{\Gamma = \{{x_1}_{\mbox{ }p_1(\mulpol{n})}:A_1,\dots,}\mathit{{x_m}_{\mbox{ }p_m(\mulpol{n})}:A_m\}}$ and $\mathit{\Phi = \{{x_1}_{\mbox{ }q_1(\mulpol{n})}:A_1,}$ $\mathit{\dots,{x_m}_{\mbox{ }q_m(\mulpol{n})}:A_m\}}$ then $\mathit{\Gamma \sqsubseteq \Phi}$ if and only if $\mathit{p_i(\mulpol{n}) \leq q_i(\mulpol{n})}$, for all $i \in \{1, ... 	,m\}$.
\end{definition}

\begin{lemma}[$\sqsubseteq$ is a Partial Order Relation]
$\sqsubseteq$ is a partial order relation between unrestricted type environments.
\end{lemma}

\subparagraph*{Type Judgments.}
A \emph{type judgment} is an expression in the form
\begin{center}
    $\judgmentpoly{\spvone}{\Gamma}{\Delta}{\Theta}{P}{z:C}$
\end{center}
where $\Gamma,\Delta,\Theta$ are the three aforementioned portions of the type 
environment, and $P$ is a process offering a session of type $C$ along the 
channel $z$. Polynomials can occur in type environments 
and types, and $\spvone$ serves to declare all variables which might occur in 
those polynomials. As in $\piDILL$, we assume that all channels and variables 
declared in 
$\Gamma$, $\Delta$, and $\Theta$ are distinct and different from $z$.

\subparagraph*{Typing Rules.}
The minimal set of typing rules for terms, whose judgments are in the form $ 
\valtd{\venvone}{\spvone}{\tmone}{\btone}$, is defined in Figure~\ref{fig:termstyprules}.

\begin{figure}[H]
	\begin{prooftree}
	\RightLabel{$[Var]$}
	\AxiomC{$\vars{\btone},\vars{\venvone}\subseteq\spvone$}
	\UnaryInfC{$\valtd{\venvone,z:\btone}{\spvone}{z}{\btone}$}
	\end{prooftree}
	
	\begin{center}
		\RightLabel{$[Bool1]$}
		\AxiomC{$\vars{\venvone}\subseteq\spvone$}
		\UnaryInfC{$\valtd{\venvone}{\spvone}{\ttrue}{\booltyp}$}
		\DisplayProof
		\quad\quad
		\RightLabel{$[Bool2]$}
		\AxiomC{$\vars{\venvone}\subseteq\spvone$}
		\UnaryInfC{$\valtd{\venvone}{\spvone}{\tfalse}{\booltyp}$}
	    	\DisplayProof
	\end{center}
	
	\begin{prooftree}
	\RightLabel{$[String]$}
	\AxiomC{$|s|\leq p$}
	\AxiomC{$\vars{\venvone},\vars{p}\subseteq\spvone$}
	\BinaryInfC{$\valtd{\venvone}{\spvone}{s}{\strtyp{p}}$}
	\end{prooftree}
	
	\begin{prooftree}
	\RightLabel{$[Fun]$}
	\AxiomC{$\typeoffun{\fsone}=\btone_1,\ldots,\btone_m\rightarrow\bttwo$}
	\AxiomC{$\valtd{\venvone}{\spvone}{\valone_i}{\btone_i\subst{n}{p}}$}
	\AxiomC{$\vars{p}\subseteq\spvone$}
	\TrinaryInfC{$\valtd{\venvone}{\spvone}{\fsone_p(\valone_1,\ldots,\valone_m)}{\bttwo\subst{n}{p}}$}
	\end{prooftree}
	\caption{Typing rules for terms}
	\label{fig:termstyprules}
\end{figure}

Typing rules for processes are in Figure~\ref{fig:processestyprules}.
\begin{figure}[H]
	\centering
	\begin{center}
		\RightLabel{[$T1L$]}
		\AxiomC{$\judgmentpoly{\spvone}{\Gamma}{\Delta}{\Theta}{P}{T}$}
		\UnaryInfC{$\judgmentpoly{\spvone}{\Gamma}{\Delta, \mbox{ } x:1}{\Theta}{P}{T}$}
		\DisplayProof
		\quad\quad
		\RightLabel{$[T1R]$}
		\AxiomC{}
		\UnaryInfC{$\judgmentpoly{\spvone}{\Gamma}{\emptyenv\mbox{ }}{\Theta}{0}{x:1}$}
	    	\DisplayProof
	\end{center}
	
	\begin{prooftree}
		\RightLabel{$[T \otimes L]$}
		\AxiomC{$\judgmentpoly{\spvone}{\Gamma}{\Delta, \mbox{ }y:A,\mbox{ } x:B}{\Theta}{P}{T}$}
		\UnaryInfC{$\judgmentpoly{\spvone}{\Gamma}{\Delta, \mbox{ } x: A \otimes B}{\Theta}{\inchannel{x}{y}.P}{T}$}
	\end{prooftree}
	
	\begin{prooftree}
		\RightLabel{$[T \otimes  R]$}
		\AxiomC{$\gammapremise$}
		\AxiomC{$\judgmentpoly{\spvone}{\Gamma_1}{\Delta}{\Theta}{P}{y:A}$}
		\AxiomC{$\judgmentpoly{\spvone}{\Gamma_2}{\Delta'}{\Theta}{Q}{x:B}$}
		\TrinaryInfC{$\judgmentpoly{\spvone}{\Gamma}{\Delta,\mbox{ } \Delta'}{\Theta}{\res{y}{\outchannel{x}{y}.(\parallel{P}{Q})}}{x: A \otimes B}$}
	\end{prooftree}
	
	\begin{prooftree}
		\RightLabel{[$T \multimap L$]}
		\AxiomC{$\gammapremise$}
		\AxiomC{$\judgmentpoly{\spvone}{\Gamma_1}{\Delta}{\Theta}{P}{y:A}$}
		\AxiomC{$\judgmentpoly{\spvone}{\Gamma_2}{\Delta', \mbox{ } x:B}{\Theta}{Q}{T}$}
		\TrinaryInfC{$\judgmentpoly{\spvone}{\Gamma}{\Delta, \mbox{ }\Delta', \mbox{ } x: A\multimap B}{\Theta}{\res{y}{\outchannel{x}{y}.(\parallel{P}				{Q})}}{T}$}
	\end{prooftree}
	
	\begin{prooftree}
		\RightLabel{$[T \multimap R]$}
		\AxiomC{$\judgmentpoly{\spvone}{\Gamma}{\Delta,\mbox{ }y:A}{\Theta}{P}{x:B}$}
		\UnaryInfC{$\judgmentpoly{\spvone}{\Gamma}{\Delta}{\Theta}{\inchannel{x}{y}.P}{x:A\multimap B}$}
	\end{prooftree}
	
	\begin{prooftree}
		\RightLabel{$[T_{cut}]$}
		\AxiomC{$\gammapremise$}
		\AxiomC{$\judgmentpoly{\spvone}{\Gamma_1}{\Delta}{\Theta}{P}{x:A}$}
		\AxiomC{$\judgmentpoly{\spvone}{\Gamma_2}{\Delta', \mbox{ }x:A}{\Theta}{Q}{T}$}
		\TrinaryInfC{$\judgmentpoly{\spvone}{\Gamma}{\Delta, \mbox{ }\Delta'}{\Theta}{\res{x}{(\parallel{P}{Q})}}{T}$}
	\end{prooftree}
	
	\begin{prooftree}
		\RightLabel{$[T_{cut^!}]$}
		\AxiomC{$p*\gammapremise$}
		\AxiomC{$\judgmentpoly{\spvone}{\Gamma_1}{\emptyenv}{\Theta}{P}{y:A}$}
		\AxiomC{$\judgmentpoly{\spvone}{\Gamma_2,u_{p}:A}{\Delta}{\Theta}{Q}{T}$}
		\TrinaryInfC{$\judgmentpoly{\spvone}{\Gamma}{\Delta}{\Theta}{\res{u}({\inrepl{u}{y}.\parallel{P}{Q}})}{T}$}
	\end{prooftree}

	\begin{prooftree}
		\RightLabel{$[T_{copy}]$}
		\AxiomC{$\judgmentpoly{\spvone}{\Gamma,\mbox{ } u_{p}:A}{\Delta,\mbox{ } y:A } {\Theta}{P}{T}$}
		\UnaryInfC{$\judgmentpoly{\spvone}{\Gamma,\mbox{ } u_{p+1}:A} {\Delta}{\Theta}{\res{y}{\outchannel{u}{y}.P}}{T}$}
	\end{prooftree}
	
	\begin{prooftree}
		\RightLabel{$[T!_{p}L]$}
		\AxiomC{$\judgmentpoly{\spvone}{\Gamma,\mbox{ } u_{p}:A } {\Delta}{\Theta}{P}{T}$}
		\UnaryInfC{$\judgmentpoly{\spvone}{\Gamma}{\Delta,\mbox{ }x:!_{p}A}{\Theta}{P\{x/u\}}{T}$}
	\end{prooftree}

	\begin{prooftree}
		\RightLabel{$[T!_{p}R]$}
		\AxiomC{$\judgmentpoly{\spvone}{\Gamma}{\emptyenv \mbox{ }} {\Theta}{Q}{y:A}$}
		\UnaryInfC{$\judgmentpoly{\spvone}{p*\Gamma}{\emptyenv\mbox{ }} {\Theta}{\inrepl{x}{y}.Q}{x:!_{p}A}$}
	\end{prooftree}

	\begin{prooftree}
		\RightLabel{$[T \oplus L]$}
		\AxiomC{$\judgmentpoly{\spvone}{\Gamma}{\Delta,\mbox{ }x:A} {\Theta}{P}{T}$}
		\AxiomC{$\judgmentpoly{\spvone}{\Gamma}{\Delta,\mbox{ }x:B} {\Theta}{Q}{T}$}
		\BinaryInfC{$\judgmentpoly{\spvone}{\Gamma}{\Delta,\mbox{ }x:A\oplus B}{\Theta}{\caseprocess{x}{P}{Q}}{T}$}
	\end{prooftree}

	\begin{prooftree}
		\RightLabel{$[T \oplus R_1]$}
		\AxiomC{$\judgmentpoly{\spvone}{\Gamma}{\Delta}{\Theta}{P}{x:A}$}
		\UnaryInfC{$\judgmentpoly{\spvone}{\Gamma}{\Delta}{\Theta}{\selectfirstprocess{x}{P}}{x:A\oplus B}$}
	\end{prooftree}

	\begin{prooftree}
		\RightLabel{$[T \oplus R_2]$}
		\AxiomC{$\judgmentpoly{\spvone}{\Gamma}{\Delta}{\Theta}{P}{x:B}$}
		\UnaryInfC{$\judgmentpoly{\spvone}{\Gamma}{\Delta}{\Theta}{\selectsecondprocess{x}{P}}{x:A\oplus B}$}
	\end{prooftree}

	\begin{prooftree}
		\RightLabel{$[T \& L_1]$}
		\AxiomC{$\judgmentpoly{\spvone}{\Gamma}{\Delta,\mbox{ } x:A} {\Theta}{P}{T}$}
		\UnaryInfC{$\judgmentpoly{\spvone}{\Gamma}{\Delta,\mbox{ } x: A\&B}{\Theta} {\selectfirstprocess{x}{P}}{T}$}
	\end{prooftree}
	
	\begin{prooftree}
		\RightLabel{$[T \& L_2]$}
		\AxiomC{$\judgmentpoly{\spvone}{\Gamma}{\Delta,\mbox{ } x:B} {\Theta}{P}{T}$}
		\UnaryInfC{$\judgmentpoly{\spvone}{\Gamma}{\Delta,\mbox{ } x: A\&B}{\Theta} {\selectsecondprocess{x}{P}}{T}$}
	\end{prooftree}

	\begin{prooftree}
		\RightLabel{$[T \& R]$}
		\AxiomC{$\judgmentpoly{\spvone}{\Gamma}{\Delta}{\Theta}{P}{x:A}$}
		\AxiomC{$\judgmentpoly{\spvone}{\Gamma}{\Delta}{\Theta}{Q}{x:B}$}
		\BinaryInfC{$\judgmentpoly{\spvone}{\Gamma}{\Delta}{\Theta}{\caseprocess{x}{P}{Q}}{x:A\&B}$}
	\end{prooftree}

\end{figure}

\begin{figure}[H]
	\centering

	\begin{prooftree}
		\RightLabel{$[T\mathbb{S}L]$}
		\AxiomC{$\judgmentpoly{\spvone}{\Gamma}{\Delta}{\Theta,\mbox{ }x:\strtyp{p}} {Q}{T}$}
		\UnaryInfC{$\judgmentpoly{\spvone}{\Gamma}{\Delta,\mbox{ }x:\strtyp{p}}{\Theta}{\interm{x}.Q}{T}$}
	\end{prooftree}

	\begin{prooftree}
		\RightLabel{$[T\mathbb{S}R]$}
		\AxiomC{$\valtd{\venvone}{\spvone}{\valone}{\strtyp{p}}$}
		\UnaryInfC{$\judgmentpoly{\spvone}{\Gamma}{\Delta}{\Theta}{\outterm{x}{\valone}}{x:\strtyp{p}}$}
	\end{prooftree}
	
	\begin{prooftree}
		\RightLabel{$[T\mathbb{B}L]$}
		\AxiomC{$\judgmentpoly{\spvone}{\Gamma}{\Delta}{\Theta,\mbox{ }x:\booltyp} {Q}{T}$}
		\UnaryInfC{$\judgmentpoly{\spvone}{\Gamma}{\Delta,\mbox{ }x:\booltyp}{\Theta}{\interm{x}.Q}{T}$}
	\end{prooftree}
	
	\begin{prooftree}
		\RightLabel{$[T\mathbb{B}R]$}
		\AxiomC{$\valtd{\venvone}{\spvone}{\valone}{\booltyp}$}
		\UnaryInfC{$\judgmentpoly{\spvone}{\Gamma}{\Delta}{\Theta}{\outterm{x}{\valone}}{x:\booltyp}$}
	\end{prooftree}
	
	\begin{prooftree}
		\RightLabel{$[T_{term\_eval}]$}
		\AxiomC{$\valtd{\venvone}{\spvone}{\tmone}{B}$}
		\AxiomC{$\judgmentpoly{\spvone}{\Gamma}{\Delta}{\Theta, \mbox{ } x:B}{P}{T}$}
		\BinaryInfC{$\judgmentpoly{\spvone}{\Gamma}{\Delta}{\Theta}{\letprocess{x}{\tmone}\mbox{ }P}{T}$}
	\end{prooftree}
	
	\begin{prooftree}
		\RightLabel{$[T_{if\_then\_else}]$}
		\AxiomC{$\valtd{\venvone}{\spvone}{v}{\booltyp}$}
		\AxiomC{$\judgmentpoly{\spvone}{\Gamma}{\Delta}{\Theta}{P}{x:A}$}
		\AxiomC{$\judgmentpoly{\spvone}{\Gamma}{\Delta}{\Theta}{Q}{x:A}$}
		\TrinaryInfC{$\judgmentpoly{\spvone}{\Gamma}{\Delta}{\Theta}{\ifprocess{v}{P}{Q}}{x:A}$}
	\end{prooftree}
	\caption{Typing rules for processes}
	\label{fig:processestyprules}
\end{figure}

\subsection{Operational Semantics}\label{sect:operationalSem}
Due to the introduction of function symbols into the calculus we obtain processes that can exchange values with different probabilities defined by the semantics of these function symbols. Consequently, the operational semantics turns out to be probabilistic and it is obtained through a complete redefinition of the operational semantic rules of the original calculus. Let us consider the definition of probability distribution discussed in Definition \ref{probDistr} and the notion of \textit{Probabilistic Transition System (PLTS)} defined as follows
\begin{definition}[Probabilistic Labeled Transition System] \label{PLTS}
A Probabilistic Labeled Transition System (PLTS) on a set of labels $A$ is a couple $PTS=(Q,\rightarrow)$ where:
\begin{itemize}
	\item Q is a nonempty set of states
	\item $\rightarrow \mbox{ } \subseteq \mbox{ } Q \mbox{ } \times\mbox{ } A  \mbox{ } \times \mbox{ } D(Q) $ is a transition relation. Given a transition $(q,\mu, \{q_1^{r_1},...,q_m^{r_m}\}) \mbox{ }\in \mbox{ }\rightarrow $:
	\begin{itemize}
		\item q is called root;
		\item $\mu \in A$ is the label of the transition;
		\item $\{q_1^{r_1},...,q_m^{r_m}\}$ is a probability distribution on the states of $Q$.
	\end{itemize}
\end{itemize}
\end{definition}
Transition labels are given by:
\begin{center}
	$\mathit{\alpha = \tau \orsintax  \overline{x\langle y \rangle} \orsintax  x(y) \orsintax  \overline{(\nu y) x\langle y \rangle} \orsintax  x.inl \orsintax  x.inr 			\orsintax  \overline{x.inl} \orsintax  \overline{x.inr} \orsintax  { x(val) \orsintax  \overline{x\langle val \rangle}}}$
\end{center}
where $x, y$ are channels, $val$ is a value and $\tau$ denotes an internal action. As in the work of Caires and Pfenning~\cite{caires2010session}, we denote by $\subject{\alpha}$ the subject of $\alpha$ (e.g. $x$ in  $\overline{(\nu y) x\langle y \rangle}$).
The transition system for the calculus is a triple $(\mathcal{P}, A, \rightarrow)$, where $\mathcal{P}$ is the set of processes of the calculus and $\rightarrow \mbox{ }\subseteq \mbox{ } \mathcal{P} \times A \times D(\mathcal{P})$ is the minimal relationship defined by inference rules in Figure \ref{fig:operationalsemrules}.

\begin{figure}[h!]
\begin{prooftree}
	\RightLabel{$(RES) \minitab with \mbox{ }y \notin fn(\alpha)$}
	\AxiomC{$P \xrightarrow{\alpha} \indexedprobdistr{Q_i}{r_i}{i\in I}$}
	\UnaryInfC{$\res{y}{P}\xrightarrow{\alpha} \indexedprobdistr{ \res{y}{Q_i}}{r_i}{i\in I}$}
\end{prooftree}

\begin{prooftree}
	\RightLabel{\textit{(PAR)\minitab with $bn(\alpha) \cap fn(R)=\emptyset$}}
	\AxiomC{$P \xrightarrow{\alpha}\indexedprobdistr{Q_i}{r_i}{i\in I}$}
	\UnaryInfC{$\parallel{P}{R}\xrightarrow{\alpha} \indexedprobdistr{\parallel{Q_i}{R}}{r_i}{i\in I}$}
\end{prooftree}

\begin{prooftree}
	\RightLabel{$(COM)$}
	\AxiomC{$P\xrightarrow{\overline{\alpha}}\dirac{P'}$}
	\AxiomC{$Q\xrightarrow{\alpha}\dirac{Q'}$}
	\BinaryInfC{$\parallel{P}{Q}\xrightarrow{\tau}\dirac{\parallel{P'}{Q'}}$}
\end{prooftree}

\begin{prooftree}
	\RightLabel{$(CLOSE)$\minitab $with \mbox{ } y \in fn(Q)$}
	\AxiomC{$P\xrightarrow{\overline{\res{y}{\outchannel{x}{y}}}}\dirac{P'}$}
	\AxiomC{$Q\xrightarrow{\inchannel{x}{y}}\dirac{Q'}$}
	\BinaryInfC{$\parallel{P}{Q}\xrightarrow{\tau}\dirac{\res{y}{\parallel{P'}{Q'}}}$}
\end{prooftree}

\begin{prooftree}
	\RightLabel{$(OPEN)$}
	\AxiomC{$P \xrightarrow{\overline{\outchannel{x}{y}}}\dirac{Q}$}
	\UnaryInfC{$\res{y}{P}\xrightarrow{\overline{\res{y}{\outchannel{x}{y}}}} \dirac{Q}$}
\end{prooftree}

\begin{center}
	\RightLabel{$(OUT)$}
	\AxiomC{}
	\UnaryInfC{$\outchannel{x}{y}.P \xrightarrow{\overline{\outchannel{x}{y}}} \dirac{P}$}
	\DisplayProof
	\quad\quad
	\RightLabel{$(IN)$}
	\AxiomC{}
	\UnaryInfC{$\inchannel{x}{y}.P \xrightarrow{\inchannel{x}{w}} \dirac{P\{w/y\}}$}
    	\DisplayProof
\end{center}

\begin{prooftree}
	\RightLabel{$(REP)$}
	\AxiomC{}
	\UnaryInfC{$\inrepl{x}{y}.P \xrightarrow{\inchannel{x}{w}} \dirac{\parallel{P\{\ w/y \}}{\inrepl{x}{y}.P}}$}
\end{prooftree}

\begin{center}
	\RightLabel{$(LOUT)$}
	\AxiomC{}
	\UnaryInfC{$\selectfirstprocess{x}{P} \xrightarrow{\overline{x.inl}} \dirac{P}$}
	\DisplayProof
	\quad\quad
	\RightLabel{$(ROUT)$}
	\AxiomC{}
	\UnaryInfC{$\selectsecondprocess{x}{P} \xrightarrow{\overline{x.inr}} \dirac{Q}$}
    	\DisplayProof
\end{center}

\begin{center}
	\RightLabel{$(LIN)$}
	\AxiomC{}
	\UnaryInfC{$\caseprocess{x}{P}{Q} \xrightarrow{x.inl} \dirac{P}$}
	\DisplayProof
	\quad\quad
	\RightLabel{$(RIN)$}
	\AxiomC{}
	\UnaryInfC{$\caseprocess{x}{P}{Q} \xrightarrow{x.inr} \dirac{Q}$}
    	\DisplayProof
\end{center}

\begin{center}
	\RightLabel{$(OUT\_value)$}
	\AxiomC{$\valone \valred \dirac{val}$}
	\UnaryInfC{$\outterm{x}{v} \xrightarrow{\overline{x \langle val \rangle}} 0$}
	\DisplayProof
	\quad\quad
	\RightLabel{$(IN\_value)$}
	\AxiomC{$\valone \valred \dirac{val}$}
	\UnaryInfC{$\interm{x}.P \xrightarrow{x (val)} \dirac{P\{val/x\}}$}
    	\DisplayProof
\end{center}

\begin{prooftree}
	\RightLabel{$(EVAL\_term)$}
	\AxiomC{$a \valred \sd{\sde{v_1}{r_1},..., \sde{v_m}{r_m}}$}
	\UnaryInfC{$\letprocess{x}{a}\mbox{ }P \xrightarrow{\tau} \sd{\sde{P\{v_1/x\}}{r_1},..., \sde{P\{v_m/x\}}{r_m}}$}
\end{prooftree}

\begin{center}
	\RightLabel{$(IF\_true)$}
	\AxiomC{}
	\UnaryInfC{$\ifprocess{\ttrue}{P}{Q} \xrightarrow{\tau} \dirac{P}$}
	\DisplayProof
	\quad\quad
	\RightLabel{$(IF\_false)$}
	\AxiomC{}
	\UnaryInfC{$\ifprocess{\tfalse}{P}{Q} \xrightarrow{\tau} \dirac{Q}$}
	\DisplayProof
\end{center}

\caption{Operational semantics rules}
\label{fig:operationalsemrules}
\end{figure}

\section{Safety and Reachability}\label{sect:soundessandpoly}
In this section we will prove some properties about the transition system 
induced by the reduction relation $\rightarrow$, as introduced in 
Section~\ref{sect:reduction}. Before delving into the details, a couple of 
remarks are in order. Although the relation $\rightarrow$ is 
defined for 
arbitrary processes, we will be concerned with the reduction of typable 
\emph{closed} processes namely those processes which can be typed under 
\emph{empty} $\Theta$ and $\mathcal{V}$. In fact, reducing processes in which 
term variables occur free does not make sense when reduction is supposed to 
model computation (as opposed to equational reasoning), like here. When 
$\Theta$ or $\mathcal{V}$ are empty, we simply omit them from the underlying 
typing judgment. Reduction being probabilistic, it is convenient to 
introduce some other reduction relation, all derived from 
$\rightarrow$:
\begin{definition}[Auxiliary Reduction Relations]
	We first of all define a relation $\mapsto$ on plain processes by 
	stipulating that
	$P\mapsto R$ iff $P\rightarrow\distrone$ and $R\in S(\distrone)$.
	We also need another reduction relation $\Rightarrow$ as the monadic 
	lifting of $\rightarrow$, thus a relation on process distributions:
	\begin{prooftree}
		\AxiomC{$R_i \rightarrow \distrtwo_i$ for every $i\in I$}
		\UnaryInfC{$\indexedprobdistr{R_i}{r_i}{i \in I} \Rightarrow
			\sum_{i \in I} \multDistr{r_i}{\distrtwo_i}$}
	\end{prooftree}
	Finally, it is convenient to put in relation any process $P$ with the
	distribution of 
	\emph{irreducible} processes to which $P$ evaluates:
	\begin{center}
	\AxiomC{$P$ is irreducible}
	\UnaryInfC{$P\Rrightarrow\dirac{P}$}
	\DisplayProof
	\qquad\qquad
	\AxiomC{$P\rightarrow\indexedprobdistr{R_i}{r_i}{i \in I}$}
	\AxiomC{$R_i \Rrightarrow \distrtwo_i$ for every $i\in I$}
	\BinaryInfC{$P\Rrightarrow\sum_{i \in I} \multDistr{r_i}{\distrtwo_i}$}
	\DisplayProof
	\end{center}
\end{definition}
The relation $\mapsto$ is perfectly sufficient to capture the qualitative 
aspects of the other reduction relations, e.g., if $P\rightarrow\distrone$ then 
for every $R\in S(\distrone)$ it holds that $P\mapsto R$. Indeed, in the 
rest of this section we will be concerned with $\mapsto$, only.

\subsection{Subject Reduction}
The property of Subject Reduction is the minimal requisite one asks to a 
type system, and says that types are preserved along reduction. In $\piDIBLL$, 
as in $\piDILL$, this property holds:
\begin{theorem}[Subject Reduction] \label{SR}
If $\judgmentclosed{\Gamma}{\Delta}{P}{z:C}$ and $P\mapsto R$, then it holds 
that $\judgmentclosed{\Gamma}{\Delta}{R}{z:C}$.
\end{theorem}
Following~\cite{caires2010session}, this property can be proved by carefully 
inspecting how $P$ can be reduced to $R$, which can happen as a result of 
either communication between two subprocesses of $P$, the evaluation of a 
term occurring inside $P$, or the firing of a conditional construction. Many 
cases have to be analysed, some of them not 
being present in $\piDILL$. When proving subject reduction, one constantly work with type derivations. As in Caires and Pfenning's paper, we will use a linear and textual notation for type derivation, called proof terms, allowing for more compact description. More details on proof terms and their use in the subject reduction theorem can be found in~\cite{caires2010session}. Subject reduction is proved by closely following the path traced
by Caires and Pfenning~\cite{caires2010session}; as a consequence, we proceed quite quickly, concentrating our
attention on the differences with their proof:
\begin{itemize}
	\item First, some cases of the subject reduction theorem in $\piDILL$ must be modified in such a way that they take into account the polynomial limitation 		introduced into the calculus. For example, the preservation lemmas related to replicated processes must be modified as follows:
	\begin{lemma}
	Assume
	\begin{enumerate}	
	    \item $\judgmentclosedcorrispondence{\Gamma_1}{\cdot}{D}{P}{y:A}$
	    \item $\judgmentclosedcorrispondence{\Gamma_2,\mbox{ } u_{p}:A}{\Delta}{E}{Q}{z:C}$ with $Q \xrightarrow[]{\overline{\res{u}			     		
	    {\outchannel{u}{y}}}} \{Q'^{\mbox{ }1}\}$
	\end{enumerate}
	Then
	\begin{enumerate}
	    \item $\prooftermsmanipulation{\expcut{p}{D}{u}{E}}{F}$ for some $F$
	    \item $\judgmentclosedcorrispondence{\Gamma}{\Delta}{F}{R}{z:C}$ with $p*\gammapremise$ for some\\ $R \equiv \parallel{\res{u}		    {\inrepl{u}{x}}.P}{\res{y}{(\parallel{P\{y/x\}}{Q'})}}$ 
	\end{enumerate}
	\end{lemma}

	\begin{lemma}
	Assume
	\begin{enumerate}
	    \item $\judgmentclosedcorrispondence{\Gamma_1}{\cdot}{D}{P}{x:A}$
	    \item $\judgmentclosedcorrispondence{\Gamma_2,\mbox{ } u_{p}:A}{\Delta}{E}{Q}{z:C}$ with $Q \xrightarrow[]{\overline{\res{u}			    
	    {\outchannel{u}{y}}}} \{Q'^{\mbox{ }1}\}$
	\end{enumerate}
	Then
	\begin{enumerate}
	    \item $\prooftermsmanipulation{\expcut{p}{D}{u}{E}}{\expcut{r}{D}{u}{F}}$ for some $F$
	    \item $\judgmentclosedcorrispondence{\Gamma_2, \mbox{ } u_{r}:A}{\Delta}{F}{R}{z:C}$ for some $R\equiv \res{y}{(\parallel{P\{y/x\}}			     	    {Q'})}$
	\end{enumerate}
	with  $r=p$ o $r=p-1$.
	\end{lemma}
	
	\item Secondly, it is necessary to prove a preservation lemma relating to the action of input and output of a value which is defined as follows
	\begin{lemma}
	Assume
	\begin{enumerate}
	    \item $\judgmentclosedcorrispondence{\Gamma_1}{\Delta_1}{D}{P}{x:G}$ with $P\xrightarrow[]{\overline{\outchannel{x}{val}}} 						    \dirac{P'} $
	    \item $\judgmentclosedcorrispondence{\Gamma_2}{\Delta_2,\mbox{ } x:G}{E}{Q}{z:C}$ with $Q \xrightarrow[]{\inchannel{x}{val}} 
		\dirac{Q'\{val /x\}}$
	\end{enumerate}
	Then
	\begin{enumerate}
	    \item $\prooftermsmanipulation{\cut{D}{x}{E}}{F}$ for some $F$
	    \item $\judgmentclosedcorrispondence{\Gamma}{\Delta_1, \mbox{ }\Delta_2}{F}{R}{z:C}$ with $\gammapremise$ for some \\$R\equiv 		     \res{x}{(\parallel{P'}{Q'\{val /x\}})}$ 
	\end{enumerate}
	with $G::=\booltyp\mbox{ }\mid\mbox{ }\strtyp{p}$
	\end{lemma}

	\item Finally, in the proof by induction of the Theorem \ref{SR} it is also necessary to consider the cases relating to the term's evaluation process and the 		conditional process.
\end{itemize}

\subsection{Progress}
The type system $\piDIBLL$ also enforces a global progress property. Following~\cite{caires2010session} we define a $live$ function for any process $P$ as follows
\begin{definition}
For any process $P$
$$live(P) \quad iif \quad P \equiv \res{\overline{n}}{(\parallel{\pi.Q}{R})} \quad for \mbox{ } some \mbox{ }\pi.Q,R,\overline{n} $$ 
where $\pi.Q$ is a non replicated guarded process.
\end{definition}

The progress property formalized in theorem \ref{th:progress} is proved by closely following the path traced by Caires and Pfenning~\cite{caires2010session}.

\begin{theorem}[Progress]\label{th:progress}
Let $\judgmentclosedcorrispondence{\cdot}{\cdot}{D}{P}{x:1}$ then either $P$ is terminated, $P$ is a composition of replicated processes or there exists Q such that $P \mapsto Q$.
\end{theorem}
 
As in Caires and Pfenning~\cite{caires2010session}, the theorem \ref{th:progress} follows as a corollary from two auxiliary lemmas that have been modified in order to be adapted to the $\piDIBLL$ calculus. In particular, the inversion lemma that relates types with action labels is defined as follows
\begin{lemma}\label{lem:lemma1Progress}
Let $\judgmentclosedcorrispondence{\Gamma}{\Delta}{D}{P}{z:C}$. If $live(P)$ then there is a $Q$ such that either
\begin{enumerate}
	\item $P \rightarrow \distrone$ and $Q \in supp(\distrone)$, or
	\item $P \xrightarrow{\alpha} \distrone$ and $Q \in supp(\distrone)$ for $\alpha$ where $s(\alpha)\in (z, \Gamma, \Delta)$. More: if $C=!_p A$ for some $A$ then $s(\alpha) \neq z$
\end{enumerate}
\end{lemma}
Moreover, the lemma that characterizes the typing of non live processes is modified as follows
\begin{lemma}\label{lem:lemma1Progress}
Let $\judgmentclosedcorrispondence{\Gamma}{\Delta}{D}{P}{z:C}$. If not $live(P)$ then
\begin{enumerate}
	\item $C=1$ or   $C=!_p C'$ for some $p$ and $C'$
	\item $(x: A_i) \in \Delta$ implies $A_i=1$ or there is $B_i$ with $A_i= !_{p_i} B_i$
	\item $C=!_p C'$  implies $P\equiv \res{\overline{x}}{(\parallel{!z(y).R'}{R})}$
\end{enumerate}
\end{lemma}

The main differences respect to the proof given by  Caires and Pfenning concern the polynomial limitation of the exponential type and the new constructs introduced into the syntax of the calculus.

\subsection{Polytime Soundness}
\newcommand{\wgt}[1]{\mathbf{W}(#1)}
\newcommand{\size}[1]{|#1|}
As already mentioned in the Introduction, Subject Reduction is not the only 
property one is interested in proving about reduction in $\piDIBLL$. In fact, 
the latter has been designed to guarantee polynomial bounds on reduction time, 
as prescribed by the computational model of cryptography. But what do we mean 
by that, exactly? What is the underlying parameter on which the polynomial 
depends? In cryptography, computation time must be polynomial on the value of 
the so called \emph{security parameter} which, as we hinted at already, is 
modelled by an element of $\mathcal{V}$. As a consequence, what we are actually 
referring to are bounds \emph{parametric} on the value of the 
polynomial variables which $P$ mentions in its type judgments, i.e. the 
$\mathcal{V}$ in
\begin{equation}\label{equ:soundness}
\judgmentclosedterm{\spvone}{\Gamma}{\Delta}{P}{z:C}
\end{equation}
Doing so, we have to keep in mind that process reduction is only defined on 
\emph{closed} processes. We can thus proceed in three steps:
\begin{itemize}
\item
We can first of all assign a $\mathcal{V}$-polynomial $\wgt{\pi}$ to 
every 
type derivation $\pi$ with conclusion mentioning $\mathcal{V}$. This is done by 
induction on the structure of $\pi$.
\item
We then prove that for closed type derivations, $\wgt{\cdot}$ strictly 
decreases along process reduction, at the same time taking the cost of each 
reduction step into account. In other words, if $\pi$ is closed and types 
$P$ where $P\mapsto Q$, then a type derivation $\xi$ for $Q$ can be found 
such that $\wgt{\pi}\geq\wgt{\xi}+k$, where $k$ is the cost of the reduction 
leading $P$ to $Q$. (In most cases $k$ is set to be $1$, the only exception 
being the evaluation of a \texttt{let}\ operator, which might involve the 
evaluation of costly functions.)
\item
Finally, the previous two points must be proved to interact well, and this is 
done by showing that for every type derivation $\pi$ with conclusion
in the form (\ref{equ:soundness}) and for every $\spvone$-substitution $\rho$, 
there is a type derivation $\pi\rho$ with conclusion
\[
\judgmentclosed{\Gamma\rho}{\Delta\rho}{P\rho}{z:C\rho}
\]
such that, crucially, $\wgt{\pi\rho}=\wgt{\pi}\rho$. In other words, the weight 
functor on type derivations commutes well with substitutions.
\end{itemize}
Altogether, this allows us to reach the following:
\begin{theorem}[Polytime Soundness]\label{th:polysound}
For every derivation $\pi$ typing $P$ there is a polynomial $p_\pi$ such that 
for every substitution $\rho$, if $P\rho\mapsto^* Q$, then the 
overall computational cost of the aforementioned reduction is bounded by 
$p_\pi(\rho)$.  
\end{theorem}
We can thus claim that, e.g., every process $\ProcAdv$ such that
\[
\vdash^{n} \ProcAdv::c:\strtyp{p} \otimes \strtyp{p} \otimes (\strtyp{p} 
\multimap \booltyp)
\]
can actually be evaluated in probabilistic polynomial time, since out of it 
one can type the processes $R_0,R_1,R_{\mathit{fun}}$ computing the three 
components in $\ProcAdv$'s type. Moreover, since $\funsymbset$ can be made 
large enough to be complete for PPT (see, e.g.,~\cite{DLZG2014}), one can also 
claim that all probabilistic (first-order) polytime behaviours can be captured 
from within $\piDIBLL$.

\section{Typable Processes and Their Probabilistic Behaviour}\label{sect:confluence}
The properties we proved in the last section, although remarkable, are agnostic 
to the probabilistic nature of $\piDIBLL$. It is now time to investigate the 
genuinely quantitative aspects of the calculus.

\subsection{Confluence}

It might seem weird that confluence can be proved for a calculus in which 
internal probabilistic choice is available. In fact, there are \emph{two} forms 
of nondeterministic evolution $\piDIBLL$ processes can give rise to, the first 
one coming from the presence of terms which can evolve probabilistically, the 
second one instead due, 
as in $\piDILL$, to the presence of parallel composition, itself offering the 
possibility of concurrent interaction. Confluence is meant to address the 
latter, not the former. More specifically, we would like to prove that whenever 
a typable process $P$ evolves towards two distinct distributions $\distrone$ 
and $\distrtwo$, the latter can be somehow unified. 

Given two distributions of processes $\mathscr{D}=\indexedprobdistr{P_i}{r_i}{i \in I}$ and $\mathscr{E}=\indexedprobdistr{Q_j}{r_j}{j \in J}$, let us define the abbreviation  $\parallel{\mathscr{D}}{\mathscr{E}}$ for the distribution $\indexedprobdistr{\parallel{P_i}{Q_j}}{r_i \mbox{ }\cdot\mbox{ } r_j}{i \in I,\mbox{ } j \in J}$ and the abbreviation $\res{x}{\mathscr{D}}$ for the distribution $\indexedprobdistr{\res{x}{P_i}}{r_i}{i \in I}$. Furthermore, it is useful to introduce a transition relation on process distribution based on the transition relation of the PLTS (see Definition \ref{PLTS}):
 \begin{definition}[Transition Relation on Process Distribution]
 Given a distribution $\mathscr{D}=\indexedprobdistr{R_i}{r_i}{i \in I}$ and an action $\beta$, then we define $\xRightarrow[]{\beta}$ as the monadic lifting of $\xrightarrow[]{\beta}$, thus a relation on process distributions:
\begin{prooftree}
    \AxiomC{$R_i \xrightarrow[]{\beta} \mathscr{E}_i$ for every $i \in I$}
    \UnaryInfC{$\indexedprobdistr{R_i}{r_i}{i \in I} \xRightarrow[]{\beta} \sum_{i \in I} \multDistr{r_i}{\mathscr{E}_i}$}
\end{prooftree}
\end{definition}

In order to render the proof simpler, the confluence property can be defined on the operational semantics, defined in Section \ref{sect:operationalSem}, as follows:
 
 \begin{theorem}[Confluence Theorem]\label{confluence}
If $\judgmentclosed{\Gamma}{\Delta}{P}{T}$ and $\mathscr{D} \xleftarrow[]{\alpha} P \xrightarrow[]{\beta} \mathscr{E}$ then one of the following conditions hold:
\begin{enumerate}
    \item $\alpha=\beta=\tau$ and $\mathscr{D}=\mathscr{E}$
    \item $\exists a. s(\alpha)=s(\beta)=a$
    \item there is $\mathscr{F}$ such that $\mathscr{D}\xRightarrow[]{\beta} \mathscr{F} \xLeftarrow[]{\alpha} \mathscr{E}$
\end{enumerate}

\begin{proof}
By induction on the proof that $\judgmentclosed{\Gamma}{\Delta}{P}{T}$.

\begin{itemize}
    \item \underline{Case $T_{cut}$}: $P=\res{x}{(\parallel{P_1}{P_2})}$\\
    In case of parallel composition between two processes there are several sub-cases to consider:
    \begin{itemize}
        \item \textit{Sub-case $P_1 \xrightarrow[]{\alpha} \mathscr{H}_1$ and $P_1 \xrightarrow[]{\beta} \mathscr{H}_2$}\\
        We are in case 3. By inductive hypothesis on $P_1$ there exists $\mathscr{Z}$ such that $\mathscr{H}_1 \xRightarrow[]{\beta} \mathscr{Z} \xLeftarrow[]{\alpha} \mathscr{H}_2$ and by case 1 of \cref{lemma1confluence} on $P=\res{x}{(\parallel{P_1}{P_2})}$ we have that $\res{x}{(\parallel{P_1}{P_2})} \xrightarrow[]{\alpha} \res{x}{(\parallel{\mathscr{H}_1}{P_2})}$ and $\res{x}{(\parallel{P_1}{P_2})} \xrightarrow[]{\beta} \res{x}{(\parallel{\mathscr{H}_2}{P_2})}$.
        
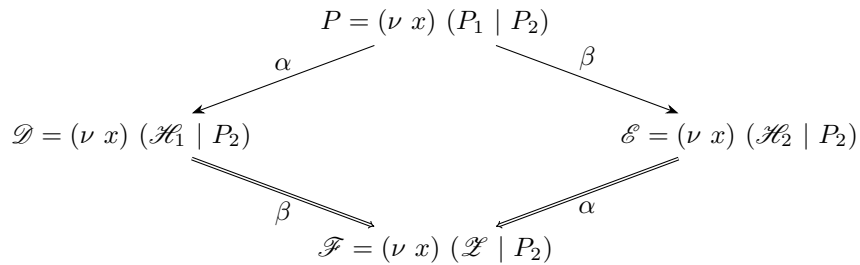
\begin{center}
            \begin{tikzpicture}
                \node (P) at (3 ,9){$P=\res{x}{(\parallel{P_1}{P_2})}$};
                \node (R) at (-1 ,7.5){$\mathscr{D}=\res{x}{(\parallel{\mathscr{H}_1}{P_2})}$};
                \node (Q) at (7 ,7.5){$\mathscr{E}=\res{x}{(\parallel{\mathscr{H}_2}{P_2})}$};
                \node (S) at (3 ,6){$\mathscr{F}=\res{x}{(\parallel{\mathscr{Z}}{P_2})}$};
                
                \draw [arrow] (P) -- node [above] {$\alpha$ } (R);
                \draw [arrow] (P) -- node [above] {$\beta$} (Q);
                \draw [-{Implies},double] (R) -- node [below] {$\beta$} (S);
                \draw [-{Implies},double] (Q) -- node [below] {$\alpha$ } (S);
            \end{tikzpicture}
        \end{center}
        where  $\mathscr{D}\xRightarrow[]{\beta} \mathscr{F}$ and $\mathscr{E}\xRightarrow[]{\alpha} \mathscr{F}$ can be proved as follows
        \begin{prooftree}
            \AxiomC{$<$By inductive hypothesis$>$}
            \UnaryInfC{$\mathscr{H}_1 \xRightarrow{\beta} \mathscr{Z}$}
            \RightLabel{(PAR)}
            \UnaryInfC{$\parallel{\mathscr{H}_1}{P_2} \xRightarrow{\beta} \parallel{\mathscr{Z}}{P_2}$}
            \RightLabel{(RES)}
            \UnaryInfC{$\mathscr{D}=\res{x}{(\parallel{\mathscr{H}_1}{P_2})} \xRightarrow{\beta}\mathscr{F}=\res{x}{(\parallel{\mathscr{Z}}{P_2})}$}
        \end{prooftree}
        \vspace{2mm}
        \begin{prooftree}
            \AxiomC{$<$By inductive hypothesis$>$}
            \UnaryInfC{$\mathscr{H}_2\xRightarrow{\alpha} \mathscr{Z}$}
            \RightLabel{(PAR)}
            \UnaryInfC{$\parallel{\mathscr{H}_2}{P_2} \xRightarrow{\alpha} \parallel{\mathscr{Z}}{P_2}$}
            \RightLabel{(RES)}
            \UnaryInfC{$\mathscr{E}=\res{x}{(\parallel{\mathscr{H}_2}{P_2})} \xRightarrow{\alpha} \mathscr{F}=\res{x}{(\parallel{\mathscr{Z}}{P_2})}$}
        \end{prooftree}
        \item \textit{Sub-case $P_2 \xrightarrow[]{\alpha} \mathscr{H}_1$ and $P_2 \xrightarrow[]{\beta} \mathscr{H}_2$}\\
        Symmetric to the previous case.
        \item \textit{Sub-case $P_1 \xrightarrow[]{\alpha} \mathscr{D}'$ and $P_2 \xrightarrow[]{\beta} \mathscr{E}'$}\\
        This is the sub-case where $P_1$ and $P_2$ evolve separately and don't communicate with each other, so we are in case 3. By case 1 of \cref{lemma1confluence} on $P=\res{x}{(\parallel{P_1}{P_2})}$ we have that $\res{x}{(\parallel{P_1}{P_2})} \xrightarrow[]{\alpha}     \res{x}{(\parallel{\mathscr{D}'}{P_2})}$ and by case 2 of \cref{lemma1confluence} we have that $\res{x}{(\parallel{P_1}{P_2})} \xrightarrow[]{\beta} \res{x}{(\parallel{P_1}{\mathscr{E}'})}$.
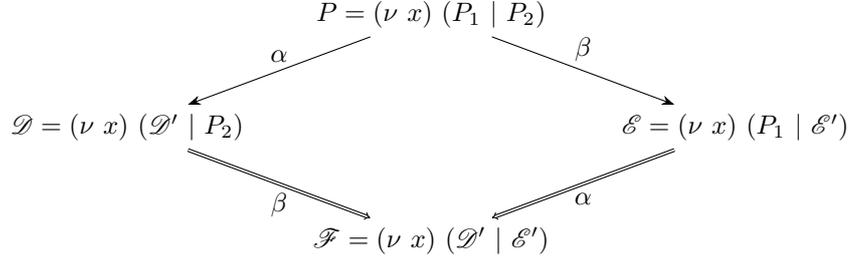
        \begin{center}
            \begin{tikzpicture}
                \node (P) at (3 ,9){$P=\res{x}{(\parallel{P_1}{P_2})}$};
                \node (R) at (-1 ,7.5){$\mathscr{D}=\res{x}{(\parallel{\mathscr{D}'}{P_2})}$};
                \node (Q) at (7 ,7.5){$\mathscr{E}=\res{x}{(\parallel{P_1}{\mathscr{E}'})}$};
                \node (S) at (3 ,6){$\mathscr{F}=\res{x}{(\parallel{\mathscr{D}'}{\mathscr{E}'})}$};
                    
                \draw [arrow] (P) -- node [above] {$\alpha$ } (R);
                \draw [arrow] (P) -- node [above] {$\beta$} (Q);
                \draw [-{Implies},double] (R) -- node [below] {$\beta$} (S);
                \draw [-{Implies},double] (Q) -- node [below] {$\alpha$ } (S);
            \end{tikzpicture}
        \end{center}
        where $\mathscr{D}\xRightarrow[]{\beta}\mathscr{F}$ and $\mathscr{E}\xRightarrow[]{\alpha}\mathscr{F}$ can be proved as follows
            \begin{prooftree}
                \AxiomC{$<$By hypothesis$>$}
                \UnaryInfC{$P_2 \xrightarrow{\beta} \mathscr{E}'$}
                \UnaryInfC{$P_2 \xRightarrow{\beta} \mathscr{E}'$}
                \RightLabel{(PAR)}
                \UnaryInfC{$\parallel{\mathscr{D}'}{P_2} \xRightarrow{\beta} \parallel{\mathscr{D}'}{\mathscr{E}'}$}
                \RightLabel{(RES)}
                \UnaryInfC{$\mathscr{D}=\res{x}{(\parallel{\mathscr{D}'}{P_2})} \xRightarrow{\beta} \mathscr{F}=\res{x}{(\parallel{\mathscr{D}'}{\mathscr{E}'})}$}
            \end{prooftree}
            \vspace{2mm}
            \begin{prooftree}
                \AxiomC{$<$By hypothesis$>$}
                \UnaryInfC{$P_1 \xrightarrow{\alpha} \mathscr{D}'$}
                \UnaryInfC{$P_1\xRightarrow{\alpha} \mathscr{D}'$}
                \RightLabel{(PAR)}
                \UnaryInfC{$\parallel{P_1}{\mathscr{E}'} \xRightarrow{\alpha} \parallel{\mathscr{D}'}{\mathscr{E}'}$}
                \RightLabel{(RES)}
                \UnaryInfC{$\mathscr{E}=\res{x}{(\parallel{P_1}{\mathscr{E}'})} \xRightarrow{\alpha} \mathscr{F}=\res{x}{(\parallel{\mathscr{D}'}{\mathscr{E}'})}$}
            \end{prooftree}
            \item \textit{Sub-case $P_1 \xrightarrow[]{\overline{\gamma}} \mathscr{D}'$ and $P_2 \xrightarrow[]{\gamma} \mathscr{E}'$}:\\
            This is the sub-case where $P_1$ and $P_2$ communicate with each other.
            \begin{itemize}
                \item If $\alpha=\beta=\tau$ and we are in case 1, then by case 3 of \cref{lemma1confluence} we have that $\mathscr{D}'=\dirac{S}$, $\mathscr{E}'=\dirac{T}$, $s(\gamma)=x$ and $\res{x}{(\parallel{P_1}{P_2})} \xrightarrow[]{\tau} \dirac{\res{x}{(\parallel{S}{T})}}=\res{x}{(\parallel{\mathscr{D}'}{\mathscr{E}'})}$, so $P_1$ and $P_2$ communicate with each other on channel $x$.
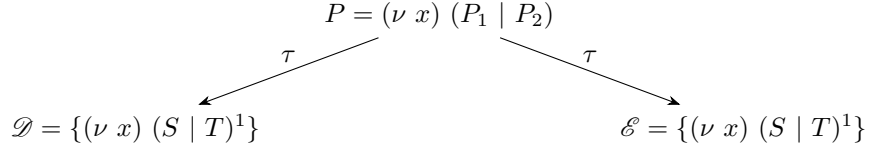
                \begin{center}
                    \begin{tikzpicture}
                        \node (P) at (3 ,9){$P=\res{x}{(\parallel{P_1}{P_2})}$};
                        \node (R) at (-1 ,7.5){$\mathscr{D}=\dirac{\res{x}{(\parallel{S}{T})}}$};
                        \node (Q) at (7 ,7.5){$\mathscr{E}=\dirac{\res{x}{(\parallel{S}{T})}}$};
                        
                        \draw [arrow] (P) -- node [above] {$\tau$ } (R);
                        \draw [arrow] (P) -- node [above] {$\tau$} (Q);
                    \end{tikzpicture}
                \end{center}
                where $\mathscr{D}=\mathscr{E}=\res{x}{(\parallel{\mathscr{D}'}{\mathscr{E}'})}$.
                \item If $\alpha=\tau$ and $\beta\neq\tau$ and we are in case 3, then by case 3 of \cref{lemma1confluence} we have that $s(\gamma)=x$ and $\res{x}{(\parallel{P_1}{P_2})} \xrightarrow[]{\tau} \res{x}{(\parallel{\mathscr{D}'}{\mathscr{E}'})}$. But $\beta\neq\tau$ so we have also that $P_2 \xrightarrow[]{\beta} \mathscr{G}$ and by case 2 of \cref{lemma1confluence} holds that $\res{x}{(\parallel{P_1}{P_2})} \xrightarrow[]{\beta} \res{x}{(\parallel{P_1}{\mathscr{G}})}$. Furthermore, by inductive hypothesis on $P_2$ we have that there exists $\mathscr{Z}$ such that $\mathscr{E}' \xRightarrow[]{\beta} \mathscr{Z} \xLeftarrow[]{\gamma} \mathscr{G}$.
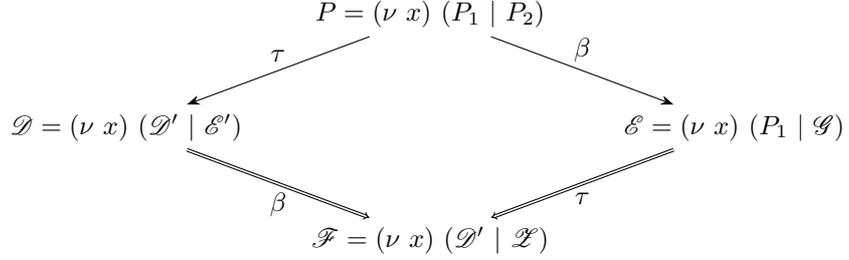
                \begin{center}
                    \begin{tikzpicture}
                        \node (P) at (3 ,9){$P=\res{x}{(\parallel{P_1}{P_2})}$};
                        \node (R) at (-1 ,7.5){$\mathscr{D}=\res{x}{(\parallel{\mathscr{D}'}{\mathscr{E}'})}$};
                        \node (Q) at (7 ,7.5){$\mathscr{E}=\res{x}{(\parallel{P_1}{\mathscr{G}})}$};
                        \node (S) at (3 ,6){$\mathscr{F}=\res{x}{(\parallel{\mathscr{D}'}{\mathscr{Z}})}$};
                            
                        \draw [arrow] (P) -- node [above] {$\tau$} (R);
                        \draw [arrow] (P) -- node [above] {$\beta$} (Q);
                        \draw [-{Implies},double] (R) -- node [below] {$\beta$} (S);
                        \draw [-{Implies},double] (Q) -- node [below] {$\tau$ } (S);
                    \end{tikzpicture}
                \end{center}
                where $\mathscr{D}\xRightarrow[]{\beta}\mathscr{F}$ and $\mathscr{E}\xRightarrow[]{\alpha}\mathscr{F}$ can be proved as follows
                \begin{prooftree}
                    \AxiomC{$<$By inductive hypothesis$>$}
                    \UnaryInfC{$\mathscr{E}' \xRightarrow{\beta} \mathscr{Z}$}
                    \RightLabel{(PAR)}
                    \UnaryInfC{$\parallel{\mathscr{D}'}{\mathscr{E}'} \xRightarrow{\beta} \parallel{\mathscr{D}'}{\mathscr{Z}}$}
                    \RightLabel{(RES)}
                    \UnaryInfC{$\mathscr{D}=\res{x}{(\parallel{\mathscr{D}'}{\mathscr{E}'})} \xRightarrow{\beta} \mathscr{F}=\res{x}{(\parallel{\mathscr{D}'}{\mathscr{Z}})}$}
                \end{prooftree}
                \vspace{3mm}
                \begin{prooftree}
                    \AxiomC{$<$By hypothesis$>$}
                    \UnaryInfC{$P_1\xrightarrow{\overline{\gamma}} \mathscr{D}'$}
                    \UnaryInfC{$P_1\xRightarrow{\overline{\gamma}} \mathscr{D}'$}
                    \AxiomC{$<$By inductive hypothesis$>$}
                    \UnaryInfC{$\mathscr{G}\xRightarrow{\gamma} \mathscr{Z}$}
                    \RightLabel{(COM)}
                    \BinaryInfC{$\parallel{P_1}{\mathscr{G}} \xRightarrow{\tau} \parallel{\mathscr{D}'}{\mathscr{Z}}$}
                    \RightLabel{(RES)}
                    \UnaryInfC{$\mathscr{E}=\res{x}{(\parallel{P_1}{\mathscr{G}})} \xRightarrow{\tau} \mathscr{F}=\res{x}{(\parallel{\mathscr{D}'}{\mathscr{Z}})}$}
                \end{prooftree}
            \item If $\beta=\tau$ and $\alpha\neq\tau$ then the proof is symmetric to the previous case.
            \end{itemize}
    \end{itemize}
    
    \item \underline{Case $T_{cut^!}$}: $P=\res{u}{(\parallel{\inrepl{u}{y}.P_1}{P_2})}$\\
    In this case we have to consider the cases in which $P_2$ contains output operations on channel $u$ (cases of interaction between $T_{cut^!}$ and $T_{copy}$), so the cases where $\inrepl{u}{y}.P_1 \xrightarrow[]{\gamma} \mathscr{D}'$ and $P_2 \xrightarrow[]{\overline{\gamma}} \mathscr{E}'$ .
    \begin{itemize}
        \item If $\alpha=\beta=\tau$ and we are in case 1, then by case 2 of \cref{lemma3confluence} on $\res{u}{(\parallel{\inrepl{u}{y}.P_1}{P_2})}$ we have that $\inrepl{u}{y}.P_1\xrightarrow[]{\gamma} \dirac{S}=\mathscr{D}'$, $P_2 \xrightarrow[]{\overline{\gamma}} \dirac{T}=\mathscr{E}'$, $s(\gamma)=u$ and \\$\res{u}{(\parallel{\inrepl{u}{y}.P_1}{P_2})} \xrightarrow[]{\tau} \dirac{\res{u}{(\parallel{S}{T})}}=\res{u}{(\parallel{\mathscr{D}'}{\mathscr{E}'})}$, so $\inrepl{u}{y}.P_1$ and $P_2$ communicate with each other on channel $u$.
        \begin{center}
            \begin{tikzpicture}
                \node (P) at (3 ,9){$P=\res{u}{(\parallel{\inrepl{u}{y}.P_1}{P_2})}$};
                \node (R) at (-1 ,7.5){$\mathscr{D}=\dirac{\res{u}{(\parallel{S}{T})}}$};
                \node (Q) at (7 ,7.5){$\mathscr{E}=\dirac{\res{u}{(\parallel{S}{T})}}$};
                        
                \draw [arrow] (P) -- node [above] {$\tau$ } (R);
                \draw [arrow] (P) -- node [above] {$\tau$} (Q);
            \end{tikzpicture}
        \end{center}
        where $\mathscr{D}=\mathscr{E}=\res{u}{(\parallel{\mathscr{D}'}{\mathscr{E}'})}$.
        \item If $\alpha=\tau$ and $\beta\neq \tau$ and we are in case 3, then by case 2 of \cref{lemma3confluence} on $\res{u}{(\parallel{\inrepl{u}{y}.P_1}{P_2})}$ we have that $\inrepl{u}{y}.P_1 \xrightarrow[]{\gamma} \dirac{S}=\mathscr{D}'$, $P_2 \xrightarrow[]{\overline{\gamma}} \dirac{T}=\mathscr{E}'$, $s(\gamma)=u$ and \\ $\res{u}{(\parallel{\inrepl{u}{y}.P_1}{P_2})} \xrightarrow[]{\tau} \dirac{\res{u}{(\parallel{S}{T})}}=\res{u}{(\parallel{\mathscr{D}'}{\mathscr{E}'})}$. But $\beta\neq\tau$ so we have also that $P_2 \xrightarrow[]{\beta} \mathscr{G}$ and by case 1 of \cref{lemma3confluence} holds that $\res{u}{(\parallel{\inrepl{u}{y}.P_1}{P_2})} \xrightarrow[]{\beta} \res{u}{(\parallel{\inrepl{u}{y}.P_1}{\mathscr{G}})}$. Furthermore, by inductive hypothesis on $P_2$ we have that there exists $\mathscr{Z}$ such that $\mathscr{E}'=\dirac{T} \xRightarrow[]{\beta} \mathscr{Z} \xLeftarrow[]{\overline{\gamma}} \mathscr{G}$.
        \begin{center}
            \begin{tikzpicture}
                \node (P) at (3 ,9){$P=\res{u}{(\parallel{\inrepl{u}{y}.P_1}{P_2})}$};
                \node (R) at (-1 ,7.5){$\mathscr{D}=\dirac{\res{u}{(\parallel{S}{T})}}=\res{u}{(\parallel{\mathscr{D}'}{\mathscr{E}'})}$};
                \node (Q) at (7 ,7.5){$\mathscr{E}=\res{u}{(\parallel{\inrepl{u}{y}.P_1}{\mathscr{G}})}$};
                \node (S) at (3 ,6){$\mathscr{F}=\res{u}{(\parallel{S}{\mathscr{Z}})}=\res{u}{(\parallel{\mathscr{D}'}{\mathscr{Z}})}$};
                    
                \draw [arrow] (P) -- node [above] {$\tau$ } (R);
                \draw [arrow] (P) -- node [above] {$\beta$} (Q);
                \draw [-{Implies},double] (R) -- node [below] {$\beta$} (S);
                \draw [-{Implies},double] (Q) -- node [below] {$\tau$ } (S);
            \end{tikzpicture}
        \end{center}
         where $\mathscr{D}\xRightarrow[]{\beta} \mathscr{F}$ and $\mathscr{E}\xRightarrow[]{\tau}\mathscr{F}$ can be proved as follows
                \begin{prooftree}
                    \AxiomC{$<$By inductive hypothesis$>$}
                    \UnaryInfC{$\mathscr{E}' \xRightarrow{\beta} \mathscr{Z}$}
                    \RightLabel{(PAR)}
                    \UnaryInfC{$\parallel{\mathscr{D}'}{\mathscr{E}'} \xRightarrow{\beta} \parallel{\mathscr{D}'}{\mathscr{Z}}$}
                    \RightLabel{(RES)}
                    \UnaryInfC{$\mathscr{D}=\res{u}{(\parallel{\mathscr{D}'}{\mathscr{E}'})} \xRightarrow{\beta} \mathscr{F}=\res{u}{(\parallel{\mathscr{D}'}{\mathscr{Z}})}$}
                \end{prooftree}
                \vspace{3mm}
                \begin{prooftree}
                    \AxiomC{$<$By hypothesis$>$}
                    \UnaryInfC{$\inrepl{u}{y}.P_1\xrightarrow{\gamma} \dirac{S}=\mathscr{D}'$}
                    \UnaryInfC{$\inrepl{u}{y}.P_1\xRightarrow{\gamma} \mathscr{D}'$}
                    \AxiomC{$<$By inductive hypothesis$>$}
                    \UnaryInfC{$\mathscr{G}\xRightarrow{\overline{\gamma}} \mathscr{Z}$}
                    \RightLabel{(COM)}
                    \BinaryInfC{$\parallel{\inrepl{u}{y}.P_1}{\mathscr{G}} \xRightarrow{\tau} \parallel{\mathscr{D}'}{\mathscr{Z}}$}
                    \RightLabel{(RES)}
                    \UnaryInfC{$\mathscr{E}=\res{u}{(\parallel{\inrepl{u}{y}.P_1}{\mathscr{G}})} \xRightarrow{\tau} S=\res{x}{(\parallel{\mathscr{D}'}{\mathscr{Z}})}$}
                \end{prooftree}
    \end{itemize}

    \item \underline{Case $T\otimes L$ and $T\multimap R$}: $P=\inchannel{x}{y}.P_1$\\
    The only action that $P$ can perform is the input action on channel $x$, so we are in case 2 where $x$ is the channel such that $s(\alpha)=s(\beta)=x$.
    
    \item \underline{Case $T\otimes R$ and $T\multimap L$}:
    $P=\res{y}{\outchannel{x}{y}.(\parallel{P_1}{P_2})}$\\
    The only action that $P$ can perform is the output action on channel $x$, $\overline{\res{y}{\outchannel{x}{y}}}$, so we are in case 2 where $x$ is the channel such that $s(\alpha)=s(\beta)=x$.
    
    \item \underline{Case $T\oplus L$ and $T\& R$}:
        $P=\caseprocess{x}{P_1}{P_2}$
    \begin{itemize}
        \item \textit{Sub-case $\alpha=\beta=x.inl$}:\\
         So we are in case 2 where $x$ is the channel such that $s(\alpha)=s(\beta)=x$.
        \item \textit{Sub-case $\alpha=\beta=x.inr$}:\\
        Equal to the previous sub-case.
    \end{itemize}
    
    \item \underline{Case $T\oplus R_1$ and $T\& L_1$}: $P=\selectfirstprocess{x}{P_1}$\\
    The only action that $P$ can perform is the selection action $\overline{x.inl}$, so we are in case 2 where $x$ is the channel such that $s(\alpha)=s(\beta)=x$.
    
    \item \underline{Case $T\oplus R_2$ and $T\& L_2$}: 
        $P=\selectsecondprocess{x}{P_1}$\\
    Equal to the previous case.
    
    \item \underline{Case $T\mathbb{S}L$ and $T\mathbb{B}L$}:
        $P=\interm{x}.P_1$\\
    The only action that $P$ can perform is the input action on channel $x$, so we are in case 2 where $x$ is the channel such that $s(\alpha)=s(\beta)=x$.
    
    \item \underline{Case $T\mathbb{S}R$ and $T\mathbb{B}R$}: $P=\outterm{x}{a}$\\
    The only action that $P$ can perform is the output action on channel $x$, so we are in case 2 where $x$ is the channel such that $s(\alpha)=s(\beta)=x$.
    
    \item \underline{Case $T!_p R$}: $P=\inrepl{u}{y}.P_1$\\
    The only action that $P$ can perform is the input action on channel $u$, so we are in case 2 where $u$ is the channel such that $s(\alpha)=s(\beta)=u$.
    
    \item \underline{Case $T_{copy}$}: $P=\res{y}{\outchannel{u}{y}.P_1}$\\
    The only action that $P$ can perform is the input action on channel $u$, so we are in case 2 where $u$ is the channel such that $s(\alpha)=s(\beta)=u$.
    
    \item \underline{Case $T_{term\_eval}$}: $P=\letprocess{x}{a}\mbox{ }P_1$\\
    The only action that $P$ can perform is $\tau$, so we are in case 1 because $\alpha=\beta=\tau$. Suppose that $[\![a]\!]=\indexedprobdistr{v_i}{r_i}{i \in I}$, so we have that:
    
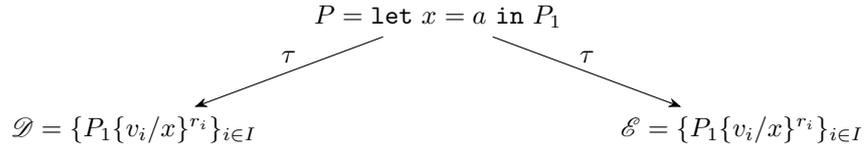
\begin{center}
        \begin{tikzpicture}
            \node (P) at (3 ,9){$P=\letprocess{x}{a}\mbox{ }P_1$};
            \node (R) at (-1 ,7.5){$\mathscr{D}=\indexedprobdistr{P_1\{v_i/x\}}{r_i}{i \in I}$};
            \node (Q) at (7 ,7.5){$\mathscr{E}=\indexedprobdistr{P_1\{v_i/x\}}{r_i}{i \in I}$};
                            
            \draw [arrow] (P) -- node [above] {$\tau$ } (R);
            \draw [arrow] (P) -- node [above] {$\tau$} (Q);
        \end{tikzpicture}
    \end{center}
    where $\mathscr{D}=\mathscr{E}$.

    \item \underline{Case $T_{if\_then\_else}$}: $P=\ifprocess{v}{P_1}{P_2}$\\
    The only action that $P$ can perform is $\tau$, so we are in case 1 because $\alpha=\beta=\tau$.
    \begin{itemize}
        \item \textit{Sub-case $[\![v]\!]=\dirac{\ttrue}$}
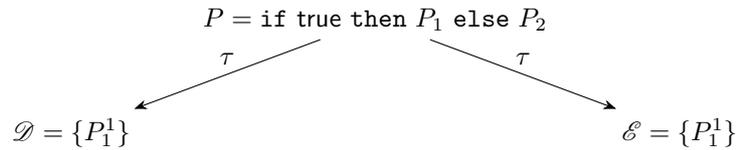
        \begin{center}
            \begin{tikzpicture}
                \node (P) at (3 ,9){$P=\ifprocess{\ttrue}{P_1}{P_2}$};
                \node (R) at (-1 ,7.5){$\mathscr{D}=\dirac{P_1}$};
                \node (Q) at (7 ,7.5){$\mathscr{E}=\dirac{P_1}$};
                            
                \draw [arrow] (P) -- node [above] {$\tau$ } (R);
                \draw [arrow] (P) -- node [above] {$\tau$} (Q);
            \end{tikzpicture}
        \end{center}
        where $\mathscr{D}=\mathscr{E}$.
        \item \textit{Sub-case $[\![v]\!]=\dirac{\tfalse}$}\\
        Symmetric to the previous sub-case.
    \end{itemize}
    
\end{itemize}
\end{proof}
\end{theorem}

Note that the cases relating to linear and exponential cut require the analysis of several sub-cases, to make the proof more readable these sub-cases have been proved in the following two lemmas:

\begin{lemma} \label{lemma1confluence}
If $\judgmentclosed{\Gamma}{\Delta_1, \mbox{ } x:A}{P}{T}$, $\judgmentclosed{\Gamma}{\Delta_2}{Q}{T}$ and $\res{x}{(\parallel{P}{Q})}\xrightarrow[]{\alpha} \mathscr{R}$, then one of the following cases holds:
\begin{enumerate}
    \item $P \xrightarrow[]{\alpha}  \mathscr{S}$ and $ \mathscr{R}=\res{x}{(\parallel{\mathscr{S}}{Q})}$
    \item $Q \xrightarrow[]{\alpha} \mathscr{S}$ and $\mathscr{R}=\res{x}{(\parallel{P}{\mathscr{S}})}$
    \item $\alpha=\tau$, $P \xrightarrow[]{\overline{\beta}}  \mathscr{S}$, $Q \xrightarrow[]{{\beta}}  \mathscr{T}$, $ \mathscr{R}=\res{x}{(\parallel{ \mathscr{S}}{ \mathscr{T}})}$ and $s(\beta)=x$
\end{enumerate}
\begin{proof}
By cases on the proof of $\res{x}{(\parallel{P}{Q})}\xrightarrow[]{\alpha} \mathscr{R}$.
\begin{itemize}
    \item \underline{Case $(PAR)$ applied to $P$}:\\
    In this case $P$ and $Q$ don't communicate with each other but $P$ performs action $\alpha$ and reduces to $\mathscr{S}$, so we are in case 1 and we can prove $\res{x}{(\parallel{P}{Q})}\xrightarrow[]{\alpha} \res{x}{(\parallel{\mathscr{S}}{Q})}$ as follows
    \begin{prooftree}
        \AxiomC{$<$Depends on $P>$}
        \UnaryInfC{$P \xrightarrow{\alpha} \mathscr{S}$}
        \RightLabel{(PAR)}
        \UnaryInfC{$\parallel{P}{Q} \xrightarrow{\alpha} \parallel{\mathscr{S}}{Q}$}
        \RightLabel{(RES)}
        \UnaryInfC{$\res{x}{(\parallel{P}{Q})}\xrightarrow{\alpha} \res{x}{(\parallel{\mathscr{S}}{Q})}$}
    \end{prooftree}
    and $s(\alpha) \neq x$ for rule $(RES)$.
    
    \item \underline{Case $(PAR)$ applied to $Q$}:\\
    We are in case 2 and the proof is symmetric to the previous case.
    
    \item \underline{Case $(COM)$}:\\
    In this case $P$ and $Q$ communicate with each other, so we are in case 3 and we can prove $\res{x}{(\parallel{P}{Q})}\xrightarrow[]{\tau} \res{x}{(\parallel{\mathscr{S}}{\mathscr{T}})}$ as follows
    \begin{prooftree}
        \AxiomC{$<$Depends on $P>$}
        \UnaryInfC{$P_1 \xrightarrow{\overline{\beta}}\mathscr{S}$}
        \AxiomC{$<$Depends on $Q>$}
        \UnaryInfC{$P_2 \xrightarrow{\beta} \mathscr{T}$}
        \RightLabel{(COM)}
        \BinaryInfC{$\parallel{P_1}{P_2} \xrightarrow{\tau} \parallel{\mathscr{S}}{\mathscr{T}}$}
        \RightLabel{(RES)}
        \UnaryInfC{$\res{x}{(\parallel{P_1}{P_2})}\xrightarrow{\tau} \res{x}{(\parallel{\mathscr{S}}{\mathscr{T}})}$}
    \end{prooftree}
    We just have prove that $s(\beta)=x$, so we have to prove that $P$ and $Q$ communicate with each other on channel $x$. This is true because, due to the restriction, we have that $x$ is the only linear channel shared between $P$ and $Q$ and consequently it represents the only channel through which $P$ and $Q$ can communicate.
\end{itemize}
\end{proof}
\end{lemma}

\begin{lemma} \label{lemma3confluence}
If $\judgmentclosed{\Gamma_1}{\emptyenv\mbox{ }}{P}{y:A}$, $\judgmentclosed{\Gamma_2, \mbox{ } u_p :A}{\Delta}{Q}{T}$ and $\res{u}{(\parallel{\inrepl{u}{y}.P}{Q})}\xrightarrow[]{\alpha} \mathscr{R}$, then one of the following cases holds:
\begin{enumerate}
    \item $Q \xrightarrow[]{\alpha} \mathscr{S}$ and $\mathscr{R}=\res{u}{(\parallel{\inrepl{u}{y}.P}{\mathscr{S}})}$
    \item $\alpha=\tau$, $\inrepl{u}{y}.P \xrightarrow[]{\beta}\mathscr{S}$, $Q\xrightarrow[]{\overline{\beta}}\mathscr{T}$, $s(\beta)=u$ and $\mathscr{R}=\res{u}{(\parallel{\mathscr{S}}{\mathscr{T}})}$ 
\end{enumerate}

\begin{proof}
By cases on the proof of $\res{u}{(\parallel{\inrepl{u}{y}.P}{Q})}\xrightarrow[]{\alpha} \mathscr{R}$.
\begin{itemize}
    \item \underline{Case $(PAR)$ applied to $Q$}:\\
    In this case $\inrepl{u}{y}.P$ and $Q$ don't communicate with each other but $Q$ performs an action $\alpha$ and reduces to $S$, so we are in case 1 and we can prove $\res{u}{(\parallel{\inrepl{u}{y}.P}{Q})}\xrightarrow[]{\alpha} \mathscr{R}$ as follows
    \begin{prooftree}
        \AxiomC{$<$Depends on $Q>$}
        \UnaryInfC{$Q \xrightarrow{\alpha} \mathscr{S}$}
        \RightLabel{(PAR)}
        \UnaryInfC{$\parallel{\inrepl{u}{y}.P}{Q} \xrightarrow{\alpha}   \mbox{ }\parallel{\inrepl{u}{y}.P}{\mathscr{S}}$}
        \RightLabel{(RES)}
        \UnaryInfC{$\res{u}{(\parallel{\inrepl{u}{y}.P}{Q})}\xrightarrow{\alpha} \res{u}{(\parallel{\inrepl{u}{y}.P}{\mathscr{S}})}$}
    \end{prooftree}
    and $s(\alpha) \neq x$ for rule $(RES)$.
    
    \item \underline{Case $(COM)$}:\\
    In this case $\inrepl{u}{y}.P$ and $Q$ communicate with each other, so we are in case 2. Note that the only action that $\inrepl{u}{y}.P$ can perform is the input action on channel $u$, so $\beta=\inchannel{u}{w}$. Consequently, $Q$ must necessarily contain an output action on channel $u$, $\overline{\beta}=\overline{\res{w}{\outchannel{u}{w}}}$, and $Q$ reduces to $\mathscr{S}$. If $\beta=\inchannel{u}{w}$ and $\overline{\beta}=\overline{\res{w}{\outchannel{u}{w}}}$ then $s(\beta)=u$, so $\inrepl{u}{y}.P$ and $Q$ communicate with each other on channel $u$. We can prove $\res{u}{(\parallel{\inrepl{u}{y}.P}{Q})}\xrightarrow[]{\alpha} \res{u}{(\parallel{\mathscr{T}}{\mathscr{S}})}$ as follows
    
    \begin{prooftree}
        \LeftLabel{(REP)}
        \AxiomC{}
        \UnaryInfC{$\inrepl{u}{y}.P \xrightarrow{\inchannel{u}{w}} \dirac{\parallel{P\{w/y\}}{\inrepl{u}{y}.P}} =\mathscr{T}$}
        \AxiomC{$<$Depends on $Q>$}
        \UnaryInfC{$Q \xrightarrow{\overline{\res{w}{\outchannel{u}{w}}}} \mathscr{S}$}
        \RightLabel{(COM)}
        \BinaryInfC{$\parallel{\inrepl{u}{y}.P}{Q} \xrightarrow{\alpha} \parallel{\mathscr{T}}{\mathscr{S}}$}
        \RightLabel{(RES)}
        \UnaryInfC{$\res{u}{(\parallel{\inrepl{u}{y}.P}{Q})}\xrightarrow[]{\alpha} \res{u}{(\parallel{\mathscr{T}}{\mathscr{S}})}$}
        \end{prooftree}
\end{itemize}
\end{proof}
\end{lemma}

In other words, while probabilistic evolution coming from terms is unavoidable, 
the choice of how to reduce a typable process does not matter, in the spirit of 
what happens in sequential languages like the $\lambda$-calculus. Notice, 
however, that confluence holds in a very strong sense here, i.e., case 3 of the 
Theorem~\ref{confluence} has the flavour of the so-called diamond property.

Among the corollaries of confluence, one can prove that the way a typable 
process is reduced is irrelevant as far as the resulting distribution is 
concerned:
\begin{corollary}[Strategy Irrelevance]\label{cor:irrelevance}
If $\judgmentclosed{\Gamma}{\Delta}{P}{T}$ and 
$\distrone \Lleftarrow P \Rrightarrow \distrtwo$, then $\distrone=\distrtwo$.
\end{corollary}

In order to prove Corollary \ref{cor:irrelevance} we need to introduce some auxiliary notions related to reduction trees and then demonstrate something about them. Given a process $P$, we define a reduction tree as follows
\begin{definition}[Reduction Tree]\label{def:reductiontree}
A Reduction Tree with root $P$ is either:
\begin{itemize}
    \item $P$ itself
    \item A triple $(P,\alpha,\mathscr{D})$ where $\mathscr{D}$ is a distribution of reduction trees, such that $P\xrightarrow[]{\alpha}\mathscr{E}$ and for every $R\in supp(\mathscr{E})$ there is $T_R \in supp(\mathscr{D})$ with root $R$ and $\mathscr{D}(T_R)=\mathscr{E}(R)$.
\end{itemize}
\end{definition}
\noindent Given a reduction tree $T$ we can extend it by adding a reduction tree $R$ to each leaf of $T$.

To adequately model the corollary by using the notion of reduction tree, we need to consider the reduction trees which are silent and normal, they are defined as follows
\begin{definition}[Silent Reduction Tree]
A reduction tree is silent if the only action occurring in it is $\tau$.
\end{definition}

\begin{definition}[Silent and Normal Reduction Tree]
A silent reduction tree is normal iif any leaf $P$ of it is such that $P$ cannot perform the $\tau$-action.
\end{definition}

Furthermore, we need a notion of kleene equivalence on reduction trees which is defined by induction as follows
\begin{definition}[Kleene Equivalence on Reduction Trees]
Two silent reduction trees $T$ and $R$ are kleene-equivalent if $[\![T]\!]=[\![R]\!]$, where the function $[\![\cdot]\!]$ on a silent reduction tree is defined by induction as follows
\begin{align*}
    [\![P]\!] & = \dirac{P}\\
    [\![ (P, \tau, \mathscr{E}=\indexedprobdistr{R_j}{r_j}{j \in J}) ]\!] &= \sum_{j \in J} \multDistr{\mathscr{E}(R_j)} {[\![ R_j ]\!]}
\end{align*}
\end{definition}
Note that the output of $[\![T]\!]$ is the probability distribution of leaf processes reachable from a silent reduction tree $T$.

Finally, Corollary \ref{cor:irrelevance} follows directly from the following theorem on silent and normal reduction trees
\begin{theorem}
Any two silent and normal reduction trees rooted at $P$ are kleene-equivalent.
\begin{proof}
We cannot prove the statement directly. Rather, we prove a strengthening of it, as follows
\begin{center}
\textit{For every pair of silent reduction tree $T$ and $R$ rooted at $P$, we can extend $T$ and $R$ into $Q$ and $S$ respectively, in such a way that $[\![Q]\!]$=$[\![S]\!]$. Moreover, the height of $Q$ and $S$ is the same.}
\end{center}
Let $n$ be the height of $T$ and $m$ the height of $R$. By induction on $n+m$:

\begin{itemize}
    \item \textit{Base case $n+m=0$}:\\
    If $n+m$ is equal to 0, then we have that $P$ is an inert process, $T=R=P$ and $n=m=0$. So, if we take $Q=T$ and $S=R$ then we have that $[\![Q]\!]$=$		    [\![S]\!]=\dirac{P}$ and the height of $Q$ and $S$ is the same and equal to $0$.
    
    \item \textit{Inductive case $n+m$ with $n,m > 0$}:\\
    Let us consider $T=(P, \tau, \mathscr{G})$ and $R=(P, \tau, \mathscr{H})$. In this case we have to consider two different sub-cases:
    \begin{itemize}
        \item If $P \xrightarrow[]{\tau} \mathscr{D}=\indexedprobdistr{U_i}{r_i}{i \in I}$, $T$ and $R$ are reduction trees such that for every $U_i \in         
        supp(\mathscr{D})$ there is:
        \begin{itemize}
            \item $T_{U_i} \in supp(\mathscr{G})$ and  $\mathscr{G}(T_{U_i}) = \mathscr{D}(U_i)$
            \item $R_{U_i} \in supp(\mathscr{H})$ and $\mathscr{H}(R_{U_i}) = \mathscr{D}(U_i)$
        \end{itemize}
        then we have that $\mathscr{G}=\mathscr{H}$ and $T=R$. So, we take $Q=T$ and $S=R$ and we have that $[\![Q]\!]$=$[\![S]\!]$ and the height of $Q$ 
        and $S$ is the same, both hold because $Q=S$.
        
        \item  If $\mathscr{E}=\indexedprobdistr{V_j}{r_j}{j \in J} \xleftarrow[]{\tau} P \xrightarrow[]{\tau} \mathscr{D}=\indexedprobdistr{U_i}{r_i}{i \in I}$,  \\			$T$ and $R$ are reduction trees such that: 
        \begin{itemize}
            \item for every $U_i \in \mathscr{D}$ there is a reduction tree $T_{U_i}$ such that $T_{U_i} \in \mathscr{G}$ and $\mathscr{D}(U_i) = \mathscr{G}			     (T_{U_i})$, where the height of $T_{U_i}$ is equal to $n-1$.
            \item for every $V_j \in \mathscr{E}$ there is a reduction tree $R_{V_j}$ such that $R_{V_j} \in \mathscr{H}$ and $\mathscr{E}(V_j) = \mathscr{H}
	    (R_{V_j})$, where the height of $R_{V_j}$ is equal to $m-1$.
        \end{itemize}
         By confluence we have that there exists $\mathscr{F}$ such that $\mathscr{D} \xRightarrow[]{\tau} \mathscr{F} \xLeftarrow[]{\tau} \mathscr{E}$. So, 			there exists a distribution of reduction trees $\mathscr{Z}$ such that:
         $$\mathscr{G}'= 
        \indexedprobdistr{{T'_{U_i}}}{r_i}{i \in I} \xRightarrow[]{\tau} \mathscr{Z}=\indexedprobdistr{Z_k}{r_k}{k \in K}\xLeftarrow[]{\tau}  						\indexedprobdistr{{R'_{V_j}}}{r_j}{j \in J} = \mathscr{H}'$$
        where the height of the reduction trees in $\mathscr{Z}$ are equal to $0$.
        
        Let us consider the reduction trees $T_{U_i}=(U_i, \tau, \mathscr{G})$ with height equal to $n-1$ and $T'_{U_i}=(U_i, \tau, \mathscr{Z})$ with height 			equal to $1$.
        
        Let us consider the reduction trees $R_{V_j}=(V_j, \tau, \mathscr{H})$ with height equal to $m-1$ and $R'_{V_j}=(V_j, \tau, \mathscr{Z})$ with height 			equal to $1$.
        
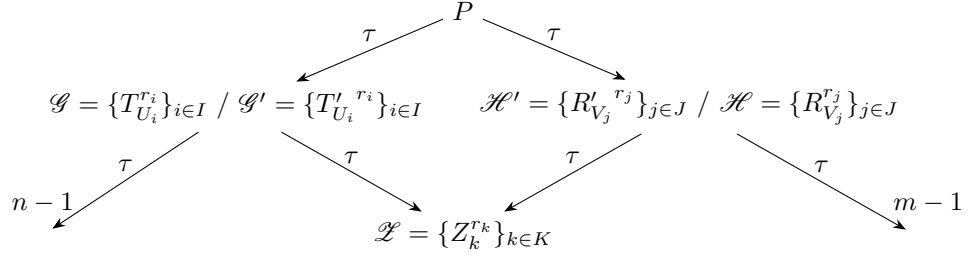
\begin{center}
            \begin{tikzpicture}[scale=0.85]
                \node (P) at (3 ,9){$P$};
                \node (G) at (-0.5 ,7.5){$\mathscr{G}= \indexedprobdistr{T_{U_i}}{r_i}{i \in I} \mbox{ }/\mbox{ } \mathscr{G}'= \indexedprobdistr{{T'_{U_i}}}{r_i}				{i \in I}$};
                \node (H) at (6.5 ,7.5){$\mathscr{H}'= \indexedprobdistr{{R'_{V_j}}}{r_j}{j \in J} \mbox{ }/\mbox{ } \mathscr{H}= \indexedprobdistr{R_{V_j}}{r_j}				{j \in J}$};
                \node (Z) at (3 ,5.5){$\mathscr{Z}=\indexedprobdistr{Z_k}{r_k}{k \in K}$};
                 
                \node (G') at (-3.5 ,5.5){};
                \node at (-3.5 ,6){$n-1$};
                
                \node (H') at (10 ,5.5){};
                \node at (10.2,6){$m-1$};
               
                \draw [arrow] (P) -- node [above] {$\tau$ } (G);
                \draw [arrow] (P) -- node [above] {$\tau$} (H);
                \draw [arrow] (G) -- node [above] {$\tau$} (Z);
                \draw [arrow] (H) -- node [above] {$\tau$} (Z);
                \draw [arrow] (G) -- node [above] {$\tau$} (G');
                \draw [arrow] (H) -- node [above] {$\tau$} (H');
            
            \end{tikzpicture}
        \end{center}
        
        By inductive hypothesis on each pair $T_{U_i}$ and $T'_{U_i}$ with $i \in I$, we can:
        \begin{itemize}
            \item Extend $T_{U_i}$ by adding a reduction tree with height $1$ to each leaf of $T_{U_i}$, obtaining $Q_{1i}$ with height $(n-1)+1$.
            \item Extend $T'_{U_i}$ by adding a reduction tree with height $n-1$ to each leaf of $T'_{U_i}$, obtaining $S_{1i}$ with height $(n-1)+1$.
        \end{itemize}
        in such a way that $[\![Q_{1i}]\!]$=$[\![S_{1i}]\!]$ and the height of $Q_{1i}$ and $S_{1i}$ is the same and equal to $(n-1)+1$.
        
        By inductive hypothesis on each pair $R_{V_j}$ and $R'_{V_j}$ with $j \in J$, we can:
        \begin{itemize}
            \item Extend $R_{V_j}$ by adding a reduction tree with height $1$ to each leaf of $R_{V_j}$, obtaining $Q_{2j}$ with height $(m-1)+1$.
            \item Extend $R'_{V_j}$ by adding a reduction tree with height $m-1$ to each leaf of $R'_{V_j}$, obtaining $S_{2j}$ with height $(m-1)+1$.
        \end{itemize}
        in such a way that $[\![Q_{2j}]\!]$=$[\![S_{2j}]\!]$ and the height of $Q_{2j}$ and $S_{2j}$ is the same and equal to $(m-1)+1$.
        
        Let us consider the reduction trees $Z_{1k}$ with height $n-1$ and $Z_{2k}$ with height $m-1$ rooted at $W_k$.
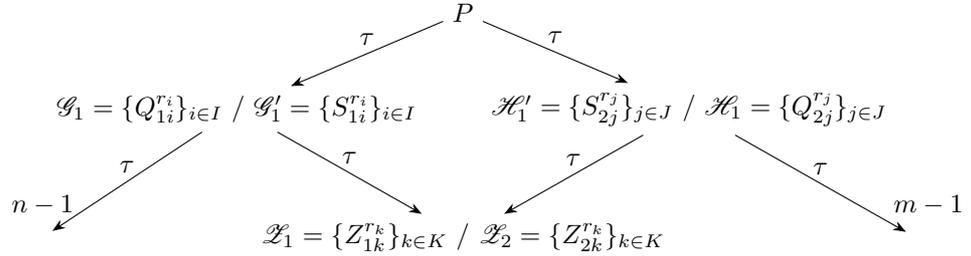
        \begin{center}
            \begin{tikzpicture}[scale=0.85]
                \node (P) at (3 ,9){$P$};
                \node (G) at (-0.5 ,7.5){$\mathscr{G}_1= \indexedprobdistr{Q_{1i}}{r_i}{i \in I} \mbox{ }/\mbox{ } \mathscr{G}'_1= \indexedprobdistr{S_{1i}}{r_i}				{i \in I}$};
                \node (H) at (6.5 ,7.5){$\mathscr{H}'_1= \indexedprobdistr{S_{2j}}{r_j}{j \in J} \mbox{ }/\mbox{ } \mathscr{H}_1= \indexedprobdistr{Q_{2j}}{r_j}				{j \in J}$};
                \node (Z) at (3 ,5.5){$\mathscr{Z}_1=\indexedprobdistr{Z_{1k}}{r_k}{k \in K} \mbox{ }/\mbox{ } \mathscr{Z}_2=\indexedprobdistr{Z_{2k}}{r_k}{k 				\in K}$};
                 
                \node (G') at (-3.5 ,5.5){};
                \node at (-3.5 ,6){$n-1$};
                \node (H') at (10 ,5.5){};
                \node at (10.2,6){$m-1$};
                        
                \draw [arrow] (P) -- node [above] {$\tau$ } (G);
                \draw [arrow] (P) -- node [above] {$\tau$} (H);
                \draw [arrow] (G) -- node [above] {$\tau$} (Z);
                \draw [arrow] (H) -- node [above] {$\tau$} (Z);
                \draw [arrow] (G) -- node [above] {$\tau$} (G');
                \draw [arrow] (H) -- node [above] {$\tau$} (H');
            \end{tikzpicture}
        \end{center}
         By inductive hypothesis on each pair $Z_{1k}$ and $Z_{2k}$ with $k \in K$, we can:
        \begin{itemize}
            \item Extend $Z_{1k}$ by adding a reduction tree with height $m-1$ to each leaf of $Z_{1k}$, obtaining $Q_{3k}$ with height $(m-1)+(n+1)$.
            \item Extend $Z_{2k}$ by adding a reduction tree with height $n-1$ to each leaf of $Z_{2k}$, obtaining $S_{3k}$ with height $(m-1)+(n+1)$.
        \end{itemize}
        in such a way that $[\![Q_{3k}]\!]$=$[\![S_{3k}]\!]$ and the height of $Q_{3k}$ and $S_{3k}$ is the same and equal to $(m-1)+(n-1)$.
        
        Let us consider the distribution of reduction trees $\mathscr{X}=\indexedprobdistr{X_{k_1}}{r_{k_1}}{k_1 \in K_1}$ which is reached by $Q_{1i}$ after 			performing $n$ $\tau$-actions, and the distribution of reduction trees $\mathscr{X}'=\indexedprobdistr{{X'_{k_1}}}{r_{k_1}}{k_1 \in K_1}$  which is 				reached by $S_{1i}$ after performing $n$ $\tau$-actions. Fixed $k_1 \in K_1$, the reduction trees in $\mathscr{X}$ and $\mathscr{X}'$ are rooted at the 			same processes because we have that $[\![Q_{1i}]\!]$=$[\![S_{1i}]\!]$ for every $i \in I$. Note that the height of $X_{k_1}$ is equal to $0$ and the height 			of $X'_{k_1}$ is equal to $m-1$.
        
        Symmetrically for the distributions reached by $Q_{2j}$ and $S_{2j}$, we have $\mathscr{Y}=\indexedprobdistr{Y_{k_2}}{r_{k_2}}{k_2 \in K_2}$ and $			\mathscr{Y'}=\indexedprobdistr{{Y'_{k_2}}}{r_{k_2}}{k_2 \in K_2}$ such that the height of $Y_{k_2}$ is equal to $0$ and the height of $Y'_{k_2}$ is equal     
	to $n-1$.
        
        
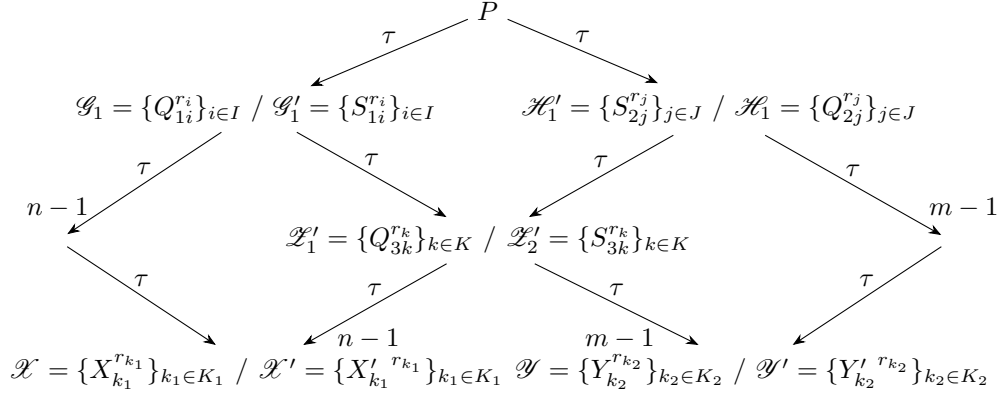
\begin{center}
            \begin{tikzpicture}[scale=0.87]
                \node (P) at (3 ,9){$P$};
                \node (G) at (-0.5 ,7.5){$\mathscr{G}_1= \indexedprobdistr{Q_{1i}}{r_i}{i \in I} \mbox{ }/\mbox{ } \mathscr{G}'_1= \indexedprobdistr{S_{1i}}{r_i}				{i \in I}$};
                \node (H) at (6.5 ,7.5){$\mathscr{H}'_1= \indexedprobdistr{S_{2j}}{r_j}{j \in J} \mbox{ }/\mbox{ } \mathscr{H}_1= \indexedprobdistr{Q_{2j}}{r_j}				{j \in J}$};
                \node (Z) at (3 ,5.5){$\mathscr{Z}'_1=\indexedprobdistr{Q_{3k}}{r_k}{k \in K} \mbox{ }/\mbox{ } \mathscr{Z}_2'=\indexedprobdistr{S_{3k}}{r_k}{k 		\in K}$};
                 
                \node (G') at (-3.5 ,5.5){};
                \node at (-3.5 ,6){$n-1$};
                \node (H') at (10 ,5.5){};
                \node at (10.2,6){$m-1$};
                
                \node (L) at (-0.5 ,3.5){$\mathscr{X}=\indexedprobdistr{X_{k_1}}{r_{k_1}}{k_1 \in K_1} \mbox{ }/\mbox{ } 						
		\mathscr{X}'=\indexedprobdistr{{X'_{k_1}}}{r_{k_1}}{k_1 \in K_1}$};
                \node at (1.2 ,4){$n-1$};
                
                \node (O) at (7 ,3.5){$\mathscr{Y}=\indexedprobdistr{Y_{k_2}}{r_{k_2}}{k_2 \in K_2} \mbox{ }/\mbox{ } \mathscr{Y'}			
		=\indexedprobdistr{{Y'_{k_2}}}{r_{k_2}}{k_2 \in K_2}$};
                \node at (5 ,4){$m-1$};
                        
                \draw [arrow] (P) -- node [above] {$\tau$ } (G);
                \draw [arrow] (P) -- node [above] {$\tau$} (H);
                \draw [arrow] (G) -- node [above] {$\tau$} (Z);
                \draw [arrow] (H) -- node [above] {$\tau$} (Z);
                \draw [arrow] (G) -- node [above] {$\tau$} (G');
                \draw [arrow] (H) -- node [above] {$\tau$} (H');
                \draw [arrow] (G') -- node [above] {$\tau$} (L);
                \draw [arrow] (Z) -- node [above] {$\tau$} (L);
                \draw [arrow] (Z) -- node [above] {$\tau$} (O);
                \draw [arrow] (H') -- node [above] {$\tau$} (O);
            \end{tikzpicture}
        \end{center}
        
        By inductive hypothesis on each pair $X_{k_1}$ and $X'_{k_1}$ with $k_1 \in K_1$, we can:
        \begin{itemize}
            \item Extend $X_{k_1}$ by adding a reduction tree with height $m-1$ to each leaf of $X_{k_1}$, obtaining $Q_{4k_1}$ with height $m-1$.
            \item Extend $X'_{k_1}$ by adding a reduction tree with height $0$ to each leaf of $X'_{k_1}$, obtaining $S_{4k_1}$ with height $m-1$.
        \end{itemize}
        in such a way that $[\![Q_{4k_1}]\!]$=$[\![S_{4k_1}]\!]$ and the height of $Q_{4k_1}$ and $S_{4k_1}$ is the same and equal to $m-1$.
        
         By inductive hypothesis on each pair $Y_{k_2}$ and $Y'_{k_2}$ with $k_2 \in K_2$, we can:
        \begin{itemize}
            \item Extend $Y_{k_2}$ by adding a reduction tree with height $n-1$ to each leaf of $Y_{k_2}$, obtaining $Q_{5k_2}$ with height $n-1$.
            \item Extend $Y'_{k_2}$ by adding a reduction tree with height $0$ to each leaf of $Y'_{k_2}$, obtaining $S_{5k_2}$ with height $n-1$.
        \end{itemize}
        in such a way that $[\![Q_{5k_2}]\!]$=$[\![S_{5k_2}]\!]$ and the height of $Q_{5k_2}$ and $S_{5k_2}$ is the same and equal to $n-1$.
        
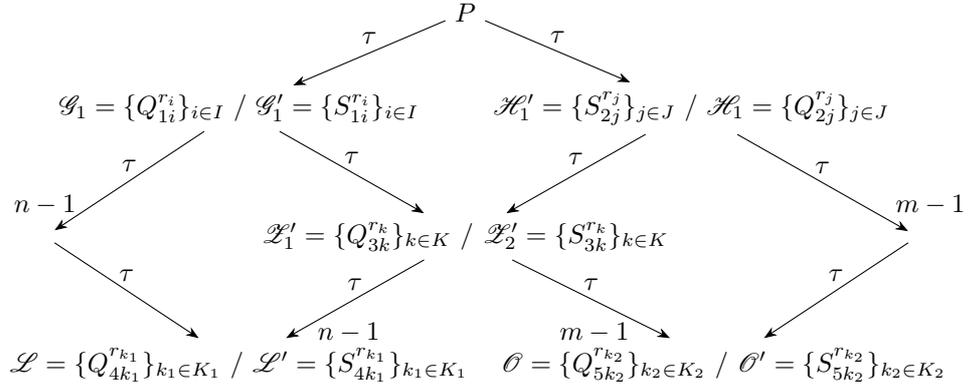
\begin{center}
            \begin{tikzpicture}[scale=0.85]
                \node (P) at (3 ,9){$P$};
                \node (G) at (-0.5 ,7.5){$\mathscr{G}_1= \indexedprobdistr{Q_{1i}}{r_i}{i \in I} \mbox{ }/\mbox{ } \mathscr{G}'_1= \indexedprobdistr{S_{1i}}{r_i}				{i \in I}$};
                \node (H) at (6.5 ,7.5){$\mathscr{H}'_1= \indexedprobdistr{S_{2j}}{r_j}{j \in J} \mbox{ }/\mbox{ } \mathscr{H}_1= \indexedprobdistr{Q_{2j}}{r_j}				{j \in J}$};
                \node (Z) at (3 ,5.5){$\mathscr{Z}'_1=\indexedprobdistr{Q_{3k}}{r_k}{k \in K} \mbox{ }/\mbox{ } \mathscr{Z}_2'=\indexedprobdistr{S_{3k}}{r_k}{k 
		\in K}$};
                 
                \node (G') at (-3.5 ,5.5){};
                \node at (-3.5 ,6){$n-1$};
                \node (H') at (10 ,5.5){};
                \node at (10.2,6){$m-1$};
                
                \node (L) at (-0.5 ,3.5){$\mathscr{L}=\indexedprobdistr{Q_{4k_1}}{r_{k_1}}{k_1 \in K_1} \mbox{ }/\mbox{ } 						
		\mathscr{L}'=\indexedprobdistr{S_{4k_1}}{r_{k_1}}{k_1 \in K_1}$};
                \node at (1.2 ,4){$n-1$};
                
                \node (O) at (7 ,3.5){$\mathscr{O}=\indexedprobdistr{Q_{5k_2}}{r_{k_2}}{k_2 \in K_2} \mbox{ }/\mbox{ } 
		\mathscr{O}'=\indexedprobdistr{S_{5k_2}}{r_{k_2}}{k_2 \in K_2}$};
                \node at (5 ,4){$m-1$};
                        
                \draw [arrow] (P) -- node [above] {$\tau$ } (G);
                \draw [arrow] (P) -- node [above] {$\tau$} (H);
                \draw [arrow] (G) -- node [above] {$\tau$} (Z);
                \draw [arrow] (H) -- node [above] {$\tau$} (Z);
                \draw [arrow] (G) -- node [above] {$\tau$} (G');
                \draw [arrow] (H) -- node [above] {$\tau$} (H');
                \draw [arrow] (G') -- node [above] {$\tau$} (L);
                \draw [arrow] (Z) -- node [above] {$\tau$} (L);
                \draw [arrow] (Z) -- node [above] {$\tau$} (O);
                \draw [arrow] (H') -- node [above] {$\tau$} (O);
            \end{tikzpicture}
        \end{center}
        By construction of $Q_{4k_1}$ and $S_{4k_1}$ we have that $Q_{4k_1}=S_{4k_1}$ for every $k_1 \in K_1$. Symmetrically, $Q_{5k_2}=S_{5k_2}$ for 			every $k_2 \in K_2$.
        
        In conclusion, we have that $Q_{4k_1}=S_{4k_1}$ can perform $m-1$ $\tau$-actions and reach $\mathscr{Q}=\indexedprobdistr{Q_{6e}}{r_e}{e \in E}$, 			and $Q_{5k_2}=S_{5k_2}$ can perform $n-1$ $\tau$-actions and reach $\mathscr{S}=\indexedprobdistr{S_{6e}}{r_e}{e \in E}$.
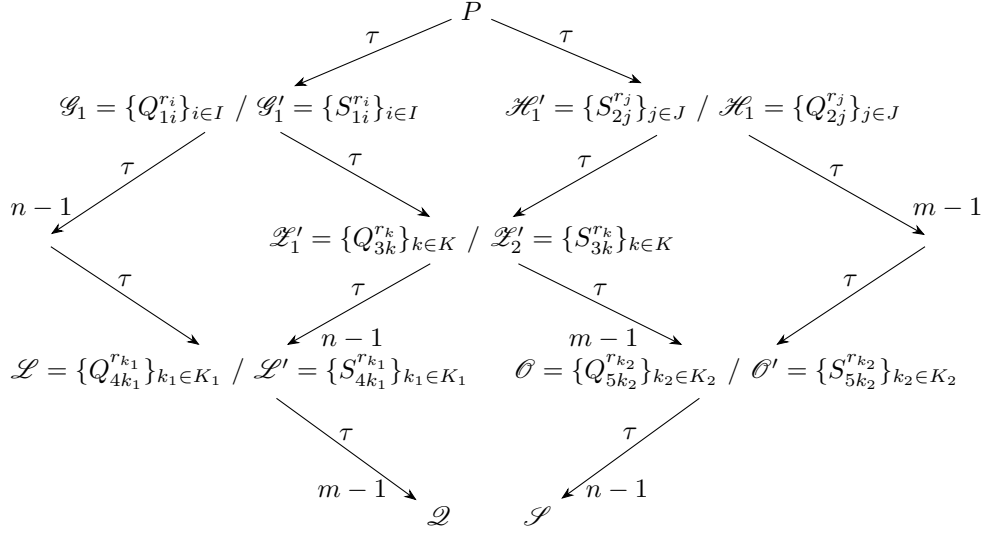
        \begin{center}
            \begin{tikzpicture}[scale=0.87]
                \node (P) at (3 ,9){$P$};
                \node (G) at (-0.5 ,7.5){$\mathscr{G}_1= \indexedprobdistr{Q_{1i}}{r_i}{i \in I} \mbox{ }/\mbox{ } \mathscr{G}'_1= \indexedprobdistr{S_{1i}}{r_i}				{i \in I}$};
                \node (H) at (6.5 ,7.5){$\mathscr{H}'_1= \indexedprobdistr{S_{2j}}{r_j}{j \in J} \mbox{ }/\mbox{ } \mathscr{H}_1= \indexedprobdistr{Q_{2j}}{r_j}				{j \in J}$};
                \node (Z) at (3 ,5.5){$\mathscr{Z}'_1=\indexedprobdistr{Q_{3k}}{r_k}{k \in K} \mbox{ }/\mbox{ } \mathscr{Z}_2'=\indexedprobdistr{S_{3k}}{r_k}{k 
		\in K}$};
                 
                \node (G') at (-3.5 ,5.5){};
                \node at (-3.5 ,6){$n-1$};
                \node (H') at (10 ,5.5){};
                \node at (10.2,6){$m-1$};
                
                \node (L) at (-0.5 ,3.5){$\mathscr{L}=\indexedprobdistr{Q_{4k_1}}{r_{k_1}}{k_1 \in K_1} \mbox{ }/\mbox{ } 				
		\mathscr{L}'=\indexedprobdistr{S_{4k_1}}{r_{k_1}}{k_1 \in K_1}$};
                \node at (1.2 ,4){$n-1$};
                
                \node (O) at (7 ,3.5){$\mathscr{O}=\indexedprobdistr{Q_{5k_2}}{r_{k_2}}{k_2 \in K_2} \mbox{ }/\mbox{ } 		
		\mathscr{O}'=\indexedprobdistr{S_{5k_2}}{r_{k_2}}{k_2 \in K_2}$};
                \node at (5 ,4){$m-1$};
                
                \node (Q) at (2.5 ,1.3){$\mathscr{Q}$};
                \node at (1.2 ,1.7){$m-1$};
                
                \node (S) at (4 ,1.3){$\mathscr{S}$};
                \node at (5.2 ,1.7){$n-1$};
                        
                \draw [arrow] (P) -- node [above] {$\tau$ } (G);
                \draw [arrow] (P) -- node [above] {$\tau$} (H);
                \draw [arrow] (G) -- node [above] {$\tau$} (Z);
                \draw [arrow] (H) -- node [above] {$\tau$} (Z);
                \draw [arrow] (G) -- node [above] {$\tau$} (G');
                \draw [arrow] (H) -- node [above] {$\tau$} (H');
                \draw [arrow] (G') -- node [above] {$\tau$} (L);
                \draw [arrow] (Z) -- node [above] {$\tau$} (L);
                \draw [arrow] (Z) -- node [above] {$\tau$} (O);
                \draw [arrow] (H') -- node [above] {$\tau$} (O);
                \draw [arrow] (L) -- node [above] {$\tau$} (Q);
                \draw [arrow] (O) -- node [above] {$\tau$} (S);
            \end{tikzpicture}
        \end{center}
        We have that the height of $Q_{6e}$ and $S_{6e}$ is the same and equal to $0$ and $[\![Q_{6e}]\!]$=$[\![S_{6e}]\!]$ because $[\![Q_{3k}]\!]$=$[\!			[S_{3k}]\!]$.
    \end{itemize}
\end{itemize}
\end{proof}
\end{theorem}

\subsection{Relational Reasoning}\label{sect:relational}
\newcommand{\obseq}{\cong}

Strategy irrelevance has a very positive impact on the definition of techniques 
for relational reasoning on typed processes. In this section, we are concerned 
precisely with introducing a form of observational equivalence, which turns out 
to have the shape one expects it to have in the realm of sequential languages. 
This is a simplification compared to similar notions from the literature 
on process algebras for the computational model of cryptography~\cite{MRST2006}.

When defining observation equivalence, we want to dub two processes as being 
equivalent when they behave \emph{the same} in \emph{any environment}. We thus 
have to 
formalize environments and observable behaviours. The former, as expected, is 
captured through the notion of a \emph{context}, which here takes the 
form of a term in which a single occurrence of the hole $\emptycontext$ is 
allowed to occur \emph{in linear position} (i.e. outside the scope of the any 
replication).
\begin{align*}
\context ::=& \emptycontext \orsintax \parallel{\context}{Q} \orsintax \parallel{P}{\context}\orsintax \res{y}\context \orsintax \outchannel{x}{y}.\context \orsintax \inchannel{x}{y}.\context \orsintax\\
& \interm{x}.\context \orsintax \selectfirstprocess{x}{\context}\orsintax \selectsecondprocess{x}{\context}\orsintax \caseprocess{x}{\context}{Q} \orsintax \caseprocess{x}{P}{\context}\orsintax\\
& \letprocess{x}{a} \mbox{ }\context \orsintax \ifprocess{v}{\context}{Q} \orsintax \ifprocess{v}{P}{\context}
\end{align*}
On top of contexts, it is routine to give a type system deriving judgments in 
the 
form
\[
\judgmentpoly{\spvone}{\Gamma}{\Delta}{\Theta}
{\contextp{(\judgmentpoly{\spvone}{\Xi}{\Psi}{\Phi}{\cdot}{T})}}{U}
\]
guaranteeing that whenever a process $P$ is such that 
$\judgmentpoly{\spvone}{\Xi}{\Psi}{\Phi}{P}{T}$, it holds that
$\judgmentpoly{\spvone}{\Gamma}{\Delta}{\Theta}{\contextp{P}}{U}$.

Talking about observations, what we are interested in observing here is the 
\emph{probability} of certain very simple events. For example, in 
Section~\ref{sect:birdeyeview} we hint at the fact that the experiment $\PrivK$ 
can be naturally modelled as a process offering a channel carrying just a 
boolean value. This will very much inform our definition, which we will now 
give. Suppose that $\judgmentobs{\spvone}{P}{x:\booltyp}$.
After having instantiated $P$ through a substitution 
$\rho:\spvone\rightarrow\NN$, some internal reduction steps turn $P\rho$ into a 
(unique) distribution $\distrone$ of irreducible processes, and the only thing 
that can happen at that time is that a boolean value $b$ is produced in output 
along the channel $x$. We write $\obspredicate{\procone}{\rho}{x}{b}$ for this 
probabilistic event, itself well defined thanks to 
Theorem~\ref{th:polysound}, Corollary~\ref{cor:irrelevance} (which witness 
the existence and unicity of $\distrone$, respectively), and Theorem~\ref{th:progress}.

Finally, we are ready to give the definition of observational equivalence.
\begin{definition}[Observational Equivalence]\label{def:obseq}
Let $\procone$ and $\proctwo$ two processes such that 
$\judgmentpoly{\spvone}{\Gamma}{\Delta}{\Theta}{\procone,\proctwo}{T}$. We say 
that $\procone$ and $\proctwo$ are observationally equivalent iff for every
context $\context$ such that 
$\judgmentobs{\spvone}
{\contextp{(\judgmentpoly{\spvone}{\Gamma}{\Delta}{\Theta}{\cdot}{T})}}{x:\booltyp}$,
there is a negligible function 
$\varepsilon:(\spvone\rightarrow\mathbb{N})\rightarrow\mathbb{R}_{[0,1]}$ such that for every $\rho$ 
it holds that
$\abs{\eventprob{\obspredicate{\contextof{\procone}}{\rho}{x}{v}}-
	 \eventprob{\obspredicate{\contextof{\proctwo}}{\rho}{x}{v}}}
 \leq\varepsilon(\rho)$.
In that case we write
$\judgmentpoly{\spvone}{\Gamma}{\Delta}{\Theta}{\procone\obseq\proctwo}{T}$,
or just $\procone\obseq\proctwo$ if this does not cause any ambiguity.
\end{definition}

Proving pairs of processes to be observationally equivalent is notoriously 
hard, due to the presence of a universal quantification over all contexts. 
However, the fact $\piDIBLL$ enjoys confluence 
facilitates the task. In particular, we can prove certain pairs of processes to 
be observationally equivalent, and this will turn out to be very useful in the 
next 
section. The first equation we can prove is the commutation of input prefixes 
and the $\mathtt{let}$ construct:
\begin{equation}\label{equ:inpvslet}
\inchannel{x}{y}.\letprocess{z}{t}\;P\obseq
\letprocess{z}{t}\;\inchannel{x}{y}.P
\end{equation}
In order to do this we have to introduce some auxiliary notions:
\begin{itemize}
    \item $\Dlet$ Distribution is the distribution obtained by fully evaluating $\letprocess{z}{t}\mbox{ } P$.
    \item $\Pletin$ stand for $\letprocess{z}{t} \mbox{ } \inchannel{x}{y}.P$.
    \item $\Pinlet$ stand for $\inchannel{x}{y}.\letprocess{z}{t}\mbox{ } P$.
    \item A $\tau$-normal process $N$ is a process that cannot perform the $\tau$-action. 
\end{itemize}

Given $\Pletin$ and $\Pinlet$ we define the following relation on process distribution:
\begin{definition}
Given two processes $\Pletin$ and $\Pinlet$, the two distributions $\distrone$ and $\distrtwo$ are $\approxdistrlet{\distrone}{\distrtwo}$ iif there are contexts $\setofcontexts,\setofcontextstwo$ such that
\begin{align*}
    \distrone &= \sd{\sde{\indexedcontext{1}{\Pletin}}{r_1},\ldots,\sde{\indexedcontext{n}{\Pletin}}{r_n}}+q_1 \cdot \mathscr{B}_1[\Dlet]+...+q_m \cdot \mathscr{B}_m[\Dlet]+\mathscr{F}\\
    \distrtwo &= \sd{\sde{\indexedcontext{1}{\Pinlet}}{r_1},\ldots,\sde{\indexedcontext{n}{\Pinlet}}{r_n}, \sde{\mathscr{B}_1[\Pinlet]}{q_1},\ldots,\sde{\mathscr{B}_m[\Pinlet]}{q_m}}+\mathscr{F}
\end{align*}
\end{definition}

Finally, the commutation of input prefixes and the let construct is proved as follows

\begin{lemma}
$\observeq{\letprocess{z}{t} \mbox{ } \inchannel{x}{y}.P}{\inchannel{x}{y}.\letprocess{z}{t}\mbox{ } P}$ where $y \notin FV(t)$
\begin{proof}\leavevmode
\begin{itemize}
    \item Given $\Pletin$ and $\Pinlet$, let $\distrone$ and $\distrtwo$ be the distributions such that $\approxdistrlet{\distrone}{\distrtwo}$.
    \item By \cref{lemmaonprocessdistr2}, we can reduce the two distributions $\distrone$ and $\distrtwo$, whose processes that compose them have bool type, in such a way that
    $$
        \distrone \xRightarrow[]{\tau}^* \mathscr{G} \quad\quad\quad 
        \distrtwo \xRightarrow[]{\tau}^* \mathscr{H} \quad\quad\quad
        \approxdistrlet{\mathscr{G}}{\mathscr{H}}
    $$
    \item By Subject Reduction Theorem (see Theorem~\ref{SR}) we know also that the reduction between processes preserves types, being the reduction between distribution based on the reduction between processes then we have that $\mathscr{G}$ and $\mathscr{H}$ are typed by bool.
    \item The distributions $\mathscr{G}$ and $\mathscr{H}$ are typed by bool so they can perform the output action on the channel with the same probability of obtaining $\ttrue$ and $\tfalse$, so we can conclude that $\observeq{\letprocess{z}{t} \mbox{ } \inchannel{x}{y}.P}{\inchannel{x}{y}.\letprocess{z}{t}\mbox{ } P}$.
\end{itemize}

\end{proof}
\end{lemma}

\begin{lemma}\label{lemmaonprocessdistr2}
Given $\Pletin$ and $\Pinlet$, let $\distrone$ and $\distrtwo$ be the distributions such that $\approxdistrlet{\distrone}{\distrtwo}$, then we can reduce $\distrone$ and $\distrtwo$ into $\mathscr{G}$ and $\mathscr{H}$ such that:
$$\approxdistrlet{\mathscr{G}}{\mathscr{H}}$$
\begin{proof}\leavevmode
\begin{itemize}
    \item By definition we have that $\distrone$ and $\distrtwo$ are in the following form:
    \begin{align*}
    \distrone &= \sd{\sde{\indexedcontext{1}{\Pletin}}{r_1},\ldots,\sde{\indexedcontext{n}{\Pletin}}{r_n}}+q_1 \cdot \mathscr{B}_1[\Dlet]+...+q_m \cdot \mathscr{B}_m[\Dlet]+\mathscr{F}\\
    \distrtwo &= \sd{\sde{\indexedcontext{1}{\Pinlet}}{r_1},\ldots,\sde{\indexedcontext{n}{\Pinlet}}{r_n}, \sde{\mathscr{B}_1[\Pinlet]}{q_1},\ldots,\sde{\mathscr{B}_m[\Pinlet]}{q_m}}+\mathscr{F}
    \end{align*}
    where $\setofcontexts,\setofcontextstwo$ are contexts.
    
    \item Let us show now that all the elements in the support of the two distributions which are not in normal form and can be evaluated (not necessarily in normal form) so as to obtain the distributions $\mathscr{G}$ and $\mathscr{H}$ that we need. Let us distinguish a few cases:
    \begin{itemize}
        \item All the elements in the support of $\mathscr{F}$ which are not in a normal form can be rewritten and we thus obtain that $\mathscr{F} \Rightarrow \mathscr{P}$ for a certain distribution $\mathscr{P}$.
        \item Now, $\indexedcontext{i}{\Pletin} \in supp (\distrone)$ cannot be $\tau$-normal and, based on where it can be reduced, we distringuish two sub-cases:
        \begin{itemize}
            \item We can have a $\tau$-reduction inside the context $\mathscr{C}_i$, and in this case we have $\mathscr{R}^i_1,...,\mathscr{R}^i_n,$\\$ r^i_n,..., r^i_n$ such that
            \begin{align*}
                \indexedcontext{i}{\Pletin} \xRightarrow{\tau} \sum^n_{j=1} r^i_j\cdot\mathscr{R}^i_j[\Pletin]\\
                \indexedcontext{i}{\Pinlet} \xRightarrow{\tau} \sum^n_{j=1} r^i_j\cdot\mathscr{R}^i_j[\Pinlet]
            \end{align*}
            
            \item The only $\tau$-reduction available in $\indexedcontext{i}{\Pletin}$ is the one in $\Pletin$, but then we can write
            \begin{align*}
                \indexedcontext{i}{\Pletin} \xRightarrow{\tau} \indexedcontext{i}{\Dlet}\\
                \indexedcontext{i}{\Pinlet} \equiv \indexedcontext{i}{\Pinlet} 
            \end{align*}
        \end{itemize}
        \item Again, $\mathscr{B}_i[\Dlet]$ cannot be $\tau$-normal but the processes in the support of $\Dlet$ are $\tau$-normal, being all inputs. Then, we can distringuish two sub-cases:
        \begin{itemize}
            \item We can have a $\tau$-reduction inside $\mathscr{B}_i$, and in this case we can proceed as above
            \item We can have that $\mathscr{B}_i$ passes a value to the processes in $\Dlet$, in this case, however, there is a distribution $\mathscr{Q}_{let}$ such that
            \begin{align*}
                \mathscr{B}_i[\Dlet] \xRightarrow{\tau} \mathscr{B}_i[\mathscr{Q}_{let}]\\
                \mathscr{B}_i[\Pinlet] \xRightarrow{\tau}^* \mathscr{B}_i[\mathscr{Q}_{let}]
            \end{align*}
            We can do that because when $\Pinlet$ receive a value as a input then it can behave exactly as $\Pletin$.
        \end{itemize}
    \end{itemize}
    \item Putting everything together, we see that $\mathscr{G}$ and $\mathscr{H}$ can indeed be found.
\end{itemize}
\end{proof}
\end{lemma}

As innocuous as it seems, the Equation \ref{equ:inpvslet} is not validated by, e.g., 
probabilistic variations on bisimilarity, being (essentially) the classic 
counterexample to the coincidence of the latter and trace equivalence.

Furthermore, an equation scheme which can be easily proved to be sound for observational 
equivalence is the one induced by so-called \emph{Kleene-equivalence}. Formally, Kleene-equivalence
is defined as follows
\begin{definition}[Kleene-equivalence]\label{def:kleeneequiv}
Two processes, $\procone$ and $\proctwo$ are kleene equivalent if $\sem{P}=\sem{Q}$, where the function $\sem{\cdot}$ on processes is defined as follows:
\begin{align*}
    \sem{N} &= \dirac{N}\\
    \sem{P} &= \sum_{j \in J} r_j \cdot \sem{R_j} \minitab if \mbox{ } P \xrightarrow{\tau} \distrtwo=\indexedprobdistr{R_j}{r_j}{j \in J}
\end{align*}
\end{definition}
In order to prove that if $P$ 
and $Q$ are Kleene equivalent then $P\obseq Q$, we have to introduce a new relation on process distributions and then proving something about it. Specifically, the relation on process distributions is defined as follows

\begin{definition}[Relation on Process Distributions] \label{def:reldistrprocess}
Given two processes $\procone$ and $\proctwo$, two distributions $\distrone$ and $\distrtwo$ are $\approxdistr{P}{Q}{\distrone}{\distrtwo}$ iif there are contexts $\setofcontexts$ and 
\begin{align*}
    \distrone &= \sd{\sde{\indexedcontext{1}{P}}{r_1},\ldots,\sde{\indexedcontext{n}{P}}{r_n}}+\mathscr{F}\\
    \distrtwo &= \sd{\sde{\indexedcontext{1}{Q}}{r_1},\ldots,\sde{\indexedcontext{n}{Q}}{r_n}}+\mathscr{F}
\end{align*}
\end{definition}
The property we want to prove on the latter relationship between process distributions is formalized as follows
\begin{lemma}\label{lemmaonprocessdistr}
If $\procone$ and $\proctwo$ are kleene equivalent and $\approxdistr{P}{Q}{\distrone}{\distrtwo}$ then $\distrone$ and $\distrtwo$ can be reduced into $\mathscr{G}$ and $\mathscr{H}$ respectively such that $\mathscr{G}$ and $\mathscr{H}$ are $\tau$-normal distributions and $\approxdistr{P}{Q}{\mathscr{G}}{\mathscr{H}}$.
\begin{proof} \leavevmode
\begin{itemize}
    \item By definition \ref{def:reldistrprocess} we have that $\distrone$ and $\distrtwo$ are distributions in the following form:
    \begin{align*}
        \distrone &= \sd{\sde{\indexedcontext{1}{P}}{r_1},\ldots,\sde{\indexedcontext{n}{P}}{r_n}}+\mathscr{F}\\
        \distrtwo &= \sd{\sde{\indexedcontext{1}{Q}}{r_1},\ldots,\sde{\indexedcontext{n}{Q}}{r_n}}+\mathscr{F}
    \end{align*}
    where $\setofcontexts$ are contexts.
    \item Let us consider the elements $\indexedcontext{i}{P}\in supp(\distrone)$ with $i\in I$ which are not $\tau$-normal. If $\{\indexedcontext{i}{P}\}_{i \in I}$ are not $\tau$-normal, then they can perform all $\tau$-actions and $\distrone \xRightarrow[]{\tau}^* \mathscr{G}$ where $\mathscr{G}$ is a $\tau$-normal process distribution  which has the following form:
    \begin{center}
        $\mathscr{G}= \indexedprobdistr{\indexedcontext{k}{P}}{r_k}{k \in \{1,\ldots,n\}-I} + \mathscr{O}$
    \end{center}
    where $\mathscr{O}$ contains $\mathscr{F}$ and all the processes to which the $\{\indexedcontext{i}{P}\}_{i \in I}$ have been reduced to.
    \item The same reasoning can be applied to $\distrtwo$, obtaining $\distrtwo \xRightarrow[]{\tau}^* \mathscr{H}$ where $\mathscr{H}$ is a $\tau$-normal process distribution which has the following form:
    \begin{center}
        $\mathscr{H}= \indexedprobdistr{\indexedcontext{h}{Q}}{r_h}{h \in \{1,\ldots,n\}-J} + \mathscr{S}$
    \end{center}
    where $\mathscr{S}$ contains $\mathscr{F}$ and all the processes to which the elements $\indexedcontext{j}{Q}\in supp(\distrtwo)$ with $j\in J$, which are not $\tau$-normal, have been reduced to.
    \item  By hypothesis we have that $\procone$ and $\proctwo$ are kleene equivalent, so we have that $\sem{P}=\sem{Q}$, this means that $\mathscr{O}=\mathscr{S}$. We have also that $\{1,\ldots,n\}-I=\{1,\ldots,n\}-J$ because the contexts in $\setofcontexts$ which are not $\tau$-normal are the same in both cases. So, by definition \ref{def:reldistrprocess}, we can conclude that $\approxdistr{P}{Q}{\mathscr{G}}{\mathscr{H}}$.
\end{itemize}
\end{proof}
\end{lemma}

Now we are ready to prove the main property as follows
\begin{lemma}
If $\procone$ and $\proctwo$ are kleene equivalent, then $\observeq{\procone}{\proctwo}$.
\begin{proof} \leavevmode
\begin{itemize}
    \item By kleene equivalence (see \ref{def:kleeneequiv}) between $\procone$ and $\proctwo$ we can say that for all contexts $\context$ we have that $\approxdistr{P}{Q}{\dirac{\contextof{P}}}{\dirac{\contextof{Q}}}$.
    \item We can reduce the two distributions $\dirac{\contextof{P}}$ and $\dirac{\contextof{Q}}$, whose processes that compose them have bool type, to a $\tau$-normal form by lemma \ref{lemmaonprocessdistr} in such a way that
    $$
        \dirac{\contextof{P}}\xRightarrow[]{\tau}^* \mathscr{G} \quad\quad\quad \dirac{\contextof{Q}}\xRightarrow[]{\tau}^* \mathscr{H} \quad\quad\quad
        \approxdistr{P}{Q}{\mathscr{G}}{\mathscr{H}}
    $$
    \item By \cref{SR} we know also that the reduction between processes preserves types, being the reduction between distribution based on the reduction between processes then we have that $\mathscr{G}$ and $\mathscr{H}$ are typed by bool.
    \item The distributions $\mathscr{G}$ and $\mathscr{H}$ are typed by bool and $\tau$-normal therefore they can only perform the output action on the channel with the same probability of obtaining $\ttrue$ and $\tfalse$, so we can conclude that $\observeq{\procone}{\proctwo}$.
\end{itemize}
\end{proof}
\end{lemma}

An equation in 
which the approximate nature of observational equivalence comes into play is 
the following one:
\[
\interm{x}.\outterm{y}{x}\obseq
\interm{x}.\letprocess{z}{\mathit{rand}}\;\letprocess{b}{\mathit{eq}(x,z)}\;
\ifprocess{b}{\outterm{y}{x}}{\outterm{y}{z}}
\]
The two involved processes,\label{key} both offering a session on $y$ having  type 
$\strtyp{n}$ by way of a session with the same type on $x$, behave the same, 
except on \emph{one} string $z$ chosen at random.

\section{A Simple Cryptographic Proof}\label{sect:cryptoproof}
\newcommand{\OUTR}{\mathit{OUTR}}
\newcommand{\OUTPR}{\mathit{OUTPR}}
\newcommand{\FAIRFLIP}{\mathit{FAIRFLIP}}
\newcommand{\ProcPrivKeyK}{\mathit{PRIVKEYK}}
\newcommand{\outc}{\mathit{out}}
\newcommand{\esprg}{\Pi_g}
\newcommand{\esotp}{\mathit{OTP}}
In this section, we put relational reasoning at work on the simple 
example we 
introduced in Section~\ref{sect:birdeyeview}. 
More specifically, we will show that the notion of observational equivalence 
from Section~\ref{sect:relational} is sufficient to prove 
Equation~\ref{eq:security} where $\sim$ is taken to be the non-contextual 
version of observational equivalence. We will do that for an encryption scheme 
$\esprg$ such that $\Enc$ is based on a pseudorandom generator 
$g$, i.e. $\Enc$ returns on input a message $m$ and a key $k$ the ciphertext 
$\mathit{xor}(m,g(k))$. When can such a function $g$ be said 
to be pseudorandom? This happens when the output of $g$ is 
indistinguishable from a truly random sequence of the same length. This, in 
turn, can be spelled out as the equation\begin{equation}\label{eq:pseudoran}
	\OUTPR_g\obseq\OUTR
\end{equation}
where $\OUTR$ is a process outputting a random string of polynomial length of a 
channel $\outc$, while $\OUTPR_g$ is a process outputting a \emph{pseudo}random 
such 
string produced according to $g$.

We now want to prove, given (\ref{eq:pseudoran}), that 
(\ref{eq:security}) holds, the latter now taking the following form:
\[
\nu\advc.(\parallel{\ProcPrivK_\Pi}{\ProcAdv})\sim\mathit{FAIRFLIP}_{\expc}
\]
Following the textbook proof of this result (see, e.g., 
~\cite{katz2020introduction}), we can structure the proof as a construction, 
out of $\ProcAdv$, of a distinguisher $D_\ProcAdv$ having type 
$\outc:\strtyp{p}\vdash^n D_\ProcAdv::\expc:\booltyp$ such that the following 
two equations hold:
\begin{align}
\nu\outc.(\parallel{\OUTPR_G}{D_\ProcAdv})&\sim
\nu\advc.(\parallel{\ProcPrivK_\Pi}{\ProcAdv})\label{eq:prg}\\
\nu\outc.(\parallel{\OUTR}{D_\ProcAdv})&\sim
\FAIRFLIP\label{eq:otp}
\end{align}
Actually, the construction of $D_\ProcAdv$ is very simple, being it the process
$\nu\advc.(\parallel{\ProcPrivKeyK_\esotp}{\ProcAdv})$, where $\esotp$ is the 
so-called one-time pad encryption scheme, and $\ProcPrivKeyK_\esotp$ 
is the process obtained from $\ProcPrivK_\esotp$ by delegating the computation 
of the key to a subprocess:
\begin{align*}
	&\ProcPrivKeyK_\esotp: && \tab \letprocess{c}{\mathit{xor}(k,m_b)}\\
	&\tab \inputfrom{m_0}{\advc} &&\tab \outputto{c}{\advc}\\
	&\tab \inputfrom{m_1}{\advc} &&\tab \inputfrom{g}{\advc}\\
	&\tab \letprocess{b}{flipcoin()} &&\tab \letprocess{r}{eq(g,b)}\\
	&\tab \inputfrom{k}{\outc} &&\tab \outputto{r}{\expc}
\end{align*}
By construction, and using some of the equations we mentioned in 
Section~\ref{sect:confluence}, one can prove that
$\ProcPrivK_{\esprg}\obseq\nu\outc.(\parallel{\OUTPR_g}{\ProcPrivKeyK_\esotp})$,
from which by congruence of $\obseq$ one derives Equation~(\ref{eq:prg}):
\begin{align*}
\nu\outc.(\parallel{\OUTPR_G}{D_\ProcAdv})&\equiv
\nu\outc.(\parallel{\OUTPR_G}{(\nu\advc.(\parallel{\ProcPrivKeyK_\Pi}{\ProcAdv}))})\\
&\equiv\nu\advc.(\parallel{\nu\outc.(\parallel{\OUTPR_G}{\ProcPrivKeyK_\esone})}{\ProcAdv})\\
&\sim\nu\advc.(\parallel{\ProcPrivK_\esone}{\ProcAdv}).
\end{align*}
Since 
$\ProcPrivK_{\esotp}\obseq\nu\outc.(\parallel{\OUTR}{\ProcPrivKeyK_\esotp})$, 
one can similarly derive that 
$$
\nu\outc.(\parallel{\OUTR_g}{D_\ProcAdv})\sim\nu\advc.(\parallel{\ProcPrivK_\esotp}{\ProcAdv}).
$$
It is well known, however, that the $\esotp$ encryption scheme is perfectly 
secure, which yields Equation~(\ref{eq:otp}).

\section{Conclusion}
\subparagraph*{Contributions.}
In this paper, we show how the discipline of session types can be useful in 
modelling and reasoning about cryptographic experiments. The use of sessions, in 
particular, allows to resolve the intrinsic nondeterminism of process algebras 
without the need for a scheduler, thus simplifying the definitional apparatus. 
The keystone to that is a confluence result, from with it follows that
the underlying reduction strategy (i.e. the scheduler) does not matter: the 
distribution of irreducible processes one obtains by reducing a typable process 
is unique. The other major technical results about the introduced system of 
session types are a polynomial bound on the time necessary to reduce any 
typable process, together with a notion of observational equivalence through 
which it is possible to 
faithfully capture computational indistinguishability, a key notion in modern cryptography.
\subparagraph*{Future Work.}
\newcommand{\ProcActAdv}{\mathit{ACTADV}}
This work, exploratory in nature, leaves many interesting problems open. Currently, 
the authors are investigating the applicability of $\piDIBLL$ 
to more complex experiments than that considered in Section 
\ref{sect:cryptoproof}. In particular, the ability to build higher-order 
sessions enables the modelling of adversaries which have access to an oracle, 
but also of experiments involving such adversaries. As an example, an \emph{active} 
adversary $\ProcActAdv$ to an encryption scheme would have type
\[
\vdash^{n} 
\ProcActAdv::c:!_{q}(\strtyp{p}\multimap\strtyp{p})\multimap\strtyp{p} \otimes 
\strtyp{p} \otimes (\strtyp{p} 
\multimap \booltyp)
\]
reflecting the availability of an oracle, modelled as a server for the 
encryption function, which can crucially be accessed only a polynomial amount 
of times. Being able to capture \emph{all} those adversaries within our 
calculus seems feasible, but requires extending the grammar 
of processes with an iterator combinator.
On the side of relational reasoning, notions of equivalence
are being studied which are sound with respect to observational 
equivalence, that is, included in it, at the same time being handier and 
avoiding any universal quantification on all contexts. The use of logical 
relations or bisimulation, already known in $\piDILL$~\cite{caires2010session} can possibly be 
adapted to $\piDIBLL$, but does not allow to faithfully capture linearity, 
falsifying equations (like~(\ref{equ:inpvslet})) which are crucial in concrete 
proofs. As a consequence, we are considering forms of trace equivalence and 
distribution-based bisimilarity~\cite{DLG2021}, since the latter are known to 
be fully 
abstract with respect to (linear) observational equivalence, in presence of effects.
\subparagraph*{Related Work.}
\newcommand{\EasyCrypt}{\texttt{EasyCrypt}}
We are certainly not the first to propose a formal calculus in 
which to model cryptographic constructions and proofs according to the 
computational model. The so called Universal Composability model (UC in the 
following), introduced by Canetti more than 
twenty years ago~\cite{Canetti2001,Canetti2020}, has been the subject of many 
investigations aimed at determining if it is possible to either simplify it or 
to capture it by way of a
calculus or process algebra (e.g.~\cite{CSV2019,LHM2019,KTR2020,BBGKS2021}). 
In  all the aforementioned works, a tension is evident between the need to be 
expressive, so as to 
capture UC proofs, and the need to keep the model simple enough, masking the 
details of probability and complexity as much as possible. $\piDIBLL$ is too 
restrictive to capture UC in its generality, but on the other hand it is very 
simple and handy. As for the approaches based on process algebras, it is once 
again worth mentioning the series of works due to Mitchell et al. and based, 
like ours, on a system of types derived from Bounded Linear 
Logic~\cite{LMMS1999,MRST2001,MRST2006}. As 
already mentioned, the main difference is the absence of a system of 
behavioural types such as session types, which forces the framework to be  
complex, relying on a  
further quantification on probabilistic schedulers, which is not needed here. 
Another very interesting line of work is the one about imperative calculi, like 
the one on which tools like \EasyCrypt\ are based~\cite{BFGSSB2014,BEGHSS2015}. 
Recently, there have been attempts at incepting some form of probabilistic 
behaviour into session types, either by allowing for probabilistic internal 
choice in multiparty sessions~\cite{AC2019}, or by 
enriching the type system itself, by making it quantitative in 
nature~\cite{IMPTT2020}. The system $\piDIBLL$ is certainly more similar to the 
former, in that randomization does not affect the type structure but only the 
process structure. This design choice is motivated by our target 
applications, namely cryptographic experiments, in which randomization 
affects \emph{which} strings protocols and adversaries produce, rather 
than their high-level behaviour. Indeed, our calculus is closer in spirit to 
some previous work on cryptographic constructions in 
$\lambda$-calculi~\cite{NZ2010} and logical systems~\cite{IK2006}, although the 
process algebraic aspects are absent there. Finally, session types have also 
been used as an handy tool guaranteeing security properties like information 
flow or access control (see, e.g.,~\cite{CCDR2010,BCDDGP15}), which are however 
different from those we are interested at here.

\bibliography{bibliography.bib}

\end{document}